\documentclass[%
aps, %
floatfix,
prb, 
twocolumn, 10pt, 
a4paper, %
superscriptaddress, %
eqsecnum, 
longbibliography,
amsfonts, amsmath, amssymb]{revtex4-2}
\usepackage[sort&compress]{natbib}
\usepackage{newtxtext}
\usepackage[varg]{newtxmath}
\usepackage{ascmac}
\usepackage{mathrsfs}
\usepackage{bm}
\usepackage{color}
\usepackage{graphicx}
\usepackage{bm}
\usepackage{braket}
\usepackage{mathrsfs}
\usepackage[normalem]{ulem}
\usepackage{hhline}
\usepackage{hyperref}
\hypersetup{%
	unicode=false, %
	bookmarksnumbered=true, %
	bookmarkstype=toc, %
	breaklinks=true, %
	colorlinks=true, %
	pdfborder={0 0 1}, %
	linkcolor=blue, %
	citecolor=blue, %
	urlcolor=blue, %
	pdftoolbar=true, %
	pdfmenubar=true, %
	pdffitwindow=false, %
	pdfstartview={FitH} %
}
\newcommand{\kB}{k_{\mathrm{B}}}

\newcommand{\eF}{\epsilon_{\mathrm{F}}}
\newcommand{\br}{\bm{r}}
\newcommand{\bk}{\bm{k}}

\newcommand{\bq}{\bm{q}}
\newcommand{\zi}{i}
\newcommand{\dd}[1]{\mathrm{d} #1\,}
\renewcommand{\Im}{\mathrm{Im}\,}
\newcommand{\tr}{\mathrm{tr}\,}

\newcommand{\R}{\mathrm{R}}
\newcommand{\A}{\mathrm{A}}
\newcommand{\X}{\mathrm{X}}
\newcommand{\Y}{\mathrm{Y}}

\begin{document}
%
\title{Microscopic theory of spin Nernst effect}
\date{\today}
\author{Junji Fujimoto}
\email[E-mail address: ]{jfujimoto@mail.saitama-u.ac.jp}
\affiliation{Department of Electrical Engineering, Electronics, and Applied Physics, Saitama University, Saitama, 338-8570, Japan}

\author{Taiki Matsushita}
\affiliation{Department of Physics, Graduate School of Science, Kyoto University, Kyoto 606-8502, Japan}
\affiliation{Yukawa Institute for Theoretical Physics, Kyoto University, Kyoto 606-8502, Japan}

\author{Masao Ogata}
\affiliation{Department of Physics, University of Tokyo, Bunkyo, Tokyo 113-0033, Japan}
\affiliation{Trans-scale Quantum Science Institute, University of Tokyo, Bunkyo-ku, Tokyo 113-0033, Japan}

\begin{abstract}
We present the microscopic theory of the spin Nernst effect, which is a transverse spin current directly induced by a temperature gradient, employing the linear response theory with Luttinger's gravitational potential method.
We consider a generic, non-interacting electron system with randomly distributed impurities and evaluate the spin current response to the gravitational potential.
Our theory takes into account a contribution of the local equilibrium current modified by Luttinger's gravitational potential and is thus consistent with the thermodynamic principle that thermal responses should vanish.
The Ward-Takahashi identities ensure that the spin Nernst current is well-behaved at low temperatures in any order of the random impurity potentials.
Furthermore, we microscopically derive the spin-current version of Mott’s formula, which associates the spin Nernst coefficient with the spin Hall conductivity.
The spin-current version of the St\v{r}eda formula is also discussed.
To demonstrate these findings, the spin Nernst current of three-dimensional Dirac electrons is computed.
Our theory is general and can therefore be extended to interacting electron systems, where Mott's formula no longer holds.
\end{abstract}
\maketitle

\section{Introduction}
Microscopic theoretical treatment of responses to a temperature gradient involves an essential difficulty because the temperature gradient is not a mechanical force but a statistical one and, therefore, cannot be written in a Hamiltonian form.
To overcome this difficulty, Luttinger introduced a fictitious gravitational potential coupled to the local energy density of the system~\cite{luttinger1964}.
He established a correspondence between the responses to the gradient of this gravitational potential and to the temperature gradient, employing Einstein's relation.
This approach using the gravitational potential has successfully described longitudinal thermal transport phenomena, including the Seebeck effect.
An analog of the gravitational potential, similar to a vector potential, was also introduced by Tatara~\cite{tatara2015}.

However, it is known that the standard calculation procedures based on the Kubo formula using the gravitational potential fail to accurately describe transverse responses to temperature gradients.
As such responses, the thermal Hall effect~\cite{smrcka1977,cooper1997,qin2011,shitade2014} and thermal spin torques~\cite{vanderbijl2014a,fujimoto2015,kohno2016} are known.
This failure is revealed when we take the low-temperature limit; these responses diverge as $T \to 0$, although the responses to temperature gradients should vanish at $T = 0$ from the thermodynamic principle.
As highlighted in the previous works~\cite{smrcka1977,cooper1997,qin2011,shitade2014,vanderbijl2014a,fujimoto2015,kohno2016}, the gravitational potential modifies the local equilibrium components, which contribute to the observable quantities as well as the non-equilibrium components evaluated from the Kubo formula.
The contributions from the local equilibrium components successfully remove the unphysical divergence at low temperatures and allow us to derive Mott's formula for the transverse responses valid at low temperatures.
Similar procedures have been found needed in magnonic systems~\cite{matsumoto2014,li2020}.
The phenomena induced by temperature gradients mentioned above have only been discussed individually, and the methodology to evaluate such linear response coefficients based on the microscopic theory with the gravitational potential, including random impurity potentials, has been absent so far.
Note that such methodology based on a semiclassical theory is discussed in Ref.~\cite{dong2020}.
We also note that a quantum kinetic theory considering the thermal vector potential is discussed~\cite{sekine2020}.

As an example of responses to a temperature gradient, we consider the spin Nernst effect, which refers to a transverse spin current directly induced by the temperature gradient without magnetic field~\cite{cheng2008,liu2010,ma2010,tauber2012,tauber2013,wimmer2013,dyrdal2016a,long2016,xiao2018,dong2020,zhang2020} and was recently observed~\cite{meyer2017,sheng2017,kim2017,bose2018}.
This effect also diverges at low temperatures when the Kubo formula is employed without considering the contribution from the local equilibrium current~\cite{ma2010}.
For a specific model, the spin Nernst effect was precisely evaluated by considering the local equilibrium spin current~\cite{dyrdal2016a,dyrdal2016}.

\begin{figure}
\centering
\includegraphics[width=\linewidth]{outline.pdf}
\caption{\label{fig:outline}Outline of the present theory.
The gravitational potential causes the following two deviations.
One is the deviation of the density matrix, which is evaluated based on the Kubo formula.
The other is the deviation of the equilibrium expectation value from the global equilibrium (into the local equilibrium), which is evaluated by the static linear response theory.
The sum of the two contributions is well-behaved at low temperatures.
Finally, by using the Einstein relation, we obtain the response to the temperature gradient.
}
\end{figure}
In this paper, we present the guiding principle of the linear response theory to the temperature gradient based on Luttinger's gravitational potential method (as outlined in Fig.~\ref{fig:outline}) and show how the contributions from the local equilibrium components successfully remove the unphysical divergence at low temperatures.
We focus on the spin Nernst effect, but our theory can be applied to other quantities induced by the temperature gradient, with the exception of the heat current.
The linear response of the spin Nernst current to the gradient of gravitational potential consists of two distinct currents: the local equilibrium spin current and the non-equilibrium spin current.
The local equilibrium current is evaluated by the canonical correlation function, and the non-equilibrium current is obtained from the Kubo formula.
We consider a generic one-particle Hamiltonian with randomly distributed impurities and evaluate the two contributions.
While the obtained non-equilibrium spin current diverges at low temperatures, the sum of the two contributions vanishes at absolute zero.
Using the Ward-Takahashi identities, we prove that the contribution from the local equilibrium spin current precisely removes the unphysical divergence of the non-equilibrium spin current at low temperatures as long as the calculation does not violate the conservation law.
The resultant spin Nernst current induced by the temperature gradient satisfies the spin-current version of Mott's formula, which associates the spin Nernst coefficient with the spin Hall conductivity.
We also discuss the spin-current version of the St\v{r}eda formula.
Finally, we compute the spin Nernst effect in the three-dimensional Dirac electron system to demonstrate our theory.
Our theory can be extended to interacting electron systems, where Mott's formula no longer holds.

The rest of this paper is organized as follows.
Section~\ref{sec:general} is devoted to the general theory of a response to a temperature gradient.
We first explain Einstein's relation and Luttinger's gravitational potential method (Sec.~\ref{sec:sub:Einsteins_relation_and_Luttingers_method}).
Then, the non-equilibrium spin current is evaluated based on the Kubo formula in the general form~(Sec.~\ref{sec:sub:response_theory}).
Section~\ref{sec:sub:evaluation} describes the evaluation of the non-equilibrium spin current and shows the unphysical divergence of the non-equilibrium spin current.
Section~\ref{sec:sub:local_equilibrium} presents the evaluation of the local equilibrium spin current and the proof of precise cancelation of the unphysical divergence based on the Ward-Takahashi identities. 
We then discuss the spin Nernst coefficient in the clean limit and the spin-current version of the St\v{r}eda formula in Sec.~\ref{sec:clean_limit_and_streda_formula}.
Finally, we show Mott's formula in Sec.~\ref{sec:sub:Mott}.
Furthermore, as a demonstration, we apply our theory to the three-dimensional Dirac electron system in Sec.~\ref{sec:3DDirac}.
Section~\ref{sec:summary} summarizes our theory.

\section{\label{sec:general}Response to temperature gradient}
In Sec.~\ref{sec:general}, we describe the linear response theory of the spin Nernst effect.
This section is inspired by Refs.~\onlinecite{vanderbijl2014a} and \onlinecite{fujimoto2015}, which discuss the response of a spin density induced by temperature gradients.
First, we review Einstein's relation and Luttinger's gravitational potential method, the foundational elements of the response theory to temperature gradients, clarifying the importance of the local equilibrium current.
Next, we review the Kubo formula for the spin current induced by the gravitational potential and evaluate the non-equilibrium spin current.
We demonstrate that the obtained non-equilibrium spin current diverges at $T = 0$, contradicting the thermodynamic principle that dictates it should vanish.
To overcome this discrepancy, we compute the additional contribution from the local equilibrium current and show that the unphysical divergence at $T = 0$ is removed when this contribution is included.
In the procedures, we show that the Ward-Takahashi identities ensure the well-behaved spin Nernst current at low temperatures.
Then, the spin-current version of the St\v{r}eda formula is discussed.
Finally, we microscopically derive the spin-current version of Mott's formula which associates the spin Nernst coefficient with the spin Hall conductivity.

\subsection{\label{sec:sub:Einsteins_relation_and_Luttingers_method}Einstein's relation and Luttinger's method}
Let us write the linear response of spin current flowing in the $i$ direction ($i = x, y, z$) with spin polarization $\alpha ( = x, y, z)$ as
\begin{align}
\langle j_{\mathrm{s}, i}^{\alpha} \rangle
	& = L_{c, ij}^{\alpha} \left( E_j + \frac{1}{e} \nabla_j \mu \right)
		+ L_{Q, ij}^{\alpha} \left( - \nabla_j \psi - \frac{\nabla_j T}{T} \right)
\label{eq:response}
,\end{align}
where $\bm{E} = (E_x, E_y, E_z)$ is the electric field, $- e$ is the electron charge, $\mu$ is the chemical potential, $\psi$ is the fictitious gravitational potential introduced by Luttinger~\cite{luttinger1964}, and $T$ is the temperature, respectively.
The repeated indices imply that the summation is to be carried out.
Equation (\ref{eq:response}) is known as Einstein's relation which indicates that the responses to the mechanical forces, $\bm{E}$ and $- {\bm \nabla} \psi$, coincide with the response to the statistical forces, $\bm{\nabla} \mu / e$ and $- \bm \nabla T/T$, respectively.
Einstein's relation allows us to formulate the response to temperature gradients, as the response to gradients of the gravitational potential.

The gravitational potential $\psi (\br)$ modifies the local energy density $Q (\br)$ as
\begin{align}
Q (\br) = h (\br) - \mu n (\br)
	\to ( h(\br) - \mu n (\br) ) ( 1 + \psi (\br) )
\label{eq:Luttinger}
,\end{align}
where $h (\br)$ and $n (\br)$ are the Hamiltonian and electron densities, respecrively.
We write the linear response to the gravitational potential as
\begin{align}
\langle j_{\mathrm{s}, i}^{\alpha} \rangle
	& = \langle j_{\mathrm{s}, i}^{\psi, \alpha} \rangle
		+ \delta \langle j_{\mathrm{s}, i}^{\alpha} \rangle
, \quad
\delta \langle j_{\mathrm{s}, i}^{\alpha} \rangle
    = \tilde{L}_{Q, ij}^{\alpha} \left( - \nabla^j \psi \right)
\label{eq:reseponse_psi}
,\end{align}
where $\langle j_{\mathrm{s}, i}^{\psi, \alpha} \rangle$ is the equilibrium expectation value of the spin current in the presence of the gravitational potential $\psi$, and $\tilde{L}_{Q, ij}^{\alpha}$ is the response coefficient to $- \nabla_j \psi$ calculated based on the Kubo formula.

We note that $\tilde{L}_{Q, ij}^{\alpha}$ may differ from $L_{Q, ij}^{\alpha}$ in Eq.~(\ref{eq:response}).
As shown below, the gravitational potential $\psi (\br)$ changes the equilibrium current $\langle j_{\mathrm{s}, i}^{\psi, \alpha} \rangle$, which gives a contribution to the response coefficient $L_{Q, ij}^{\alpha}$.

The gravitational potential modifies the physical quantities (except the quantities related to heat, such as the heat current and the heat magnetization) as $\braket{A} \to (1 + \psi (\br)) \braket{A}$ up to the linear order of $\psi (\br)$, where $\braket{A}$ is a physical quantity in the absence of the gravitational potential.
Therefore, the (local) equilibrium spin current is modified as,
\begin{align}
\langle j_{\mathrm{s}, i}^{\psi, \alpha} \rangle
	& = ( 1 + \psi (\br) ) \langle j_{\mathrm{s}, i}^{0, \alpha} (\br) \rangle
\label{eq:scaling}
.\end{align}
where $\langle j_{\mathrm{s}, i}^{0, \alpha} (\br) \rangle$ is the (local) equilibrium spin current in the absence of the gravitational potential.
As described later in Sec.~\ref{sec:sub:local_equilibrium}, the local equilibrium spin current in the absence of the gravitational potential is expressed as
\begin{align}
\langle j_{\mathrm{s}, i}^{0, \alpha} (\br) \rangle
	& = \nabla_j \mathcal{J}_{i j}^{0, \alpha} [\mu (\br), T (\br)]
\label{eq:con}
,\end{align}
where $\nabla_j \mathcal{J}_{i j}^{0, \alpha}$ is assumed to be a functional of $\mu(\br)$ and $T(\br)$. 
Using Eq.~(\ref{eq:con}), we rewrite Eq.~(\ref{eq:scaling}) as
\begin{align}
\langle j_{\mathrm{s}, i}^{\psi, \alpha} \rangle
	& = ( 1 + \psi (\br) ) \nabla_j \mathcal{J}_{i j}^{0, \alpha}
\notag \\
	& = \nabla_j \mathcal{J}_{i j}^{\psi, \alpha}
		+ \mathcal{J}_{i j}^{0, \alpha} ( - \nabla_j \psi (\br) )
\label{eq:j^psi}
\\
	& = \nabla_j \mathcal{J}_{i j}^{\psi, \alpha}
        + \langle \delta j_{\mathrm{s}, i}^{\alpha} \rangle
\notag
,\end{align}
where $\mathcal{J}_{i j}^{\psi, \alpha} = ( 1 + \psi (\br) ) \mathcal{J}_{i j}^{0, \alpha}$ is introduced as $\mathcal{J}_{i j}^{0, \alpha}$ under the grabitational potential.
The first term in Eq.~(\ref{eq:j^psi}) is the equilibrium spin current circulating the system so that it does not contribute to the $\bm q=0$ component of the spin current or $\int \langle j_{\mathrm{s}, i}^{\alpha} \rangle \dd{\br}$.
However, the second term in Eq.~(\ref{eq:j^psi}) contributes to the response coefficient $L_{Q, ij}^{\alpha}$.
Hence, we obtain the proper response coefficient; $\varDelta \langle j_{\mathrm{s}, i}^{\alpha} \rangle = L_{Q, ij}^{\alpha} (- \nabla_j \psi)$ with
\begin{align}
L_{Q, ij}^{\alpha}
	& = \tilde{L}_{Q, ij}^{\alpha} + \mathcal{J}_{i j}^{0, \alpha}
,\end{align}
where $\mathcal{J}_{i j}^{0, \alpha}$ should be evaluated at a constant $\mu(\br)$ and $T(\br)$.
We also note that $\mathcal{J}_{i j}^{0, \alpha}$ is an equilibrium quantity and thus weakly depends on the lifetime of electrons.
As shown later, $\mathcal{J}_{i j}^{0, \alpha}$ is significantly important to remove the unphysical divergence of the non-equilibrium spin current expressed by $\tilde{L}_{Q, ij}^{\alpha}$ at low temperatures.

It is worth considering the case when the spin is conserved.
In this case, the spin current can be expressed 
as $\langle \bm{j}_{\mathrm{s}}^z \rangle = \langle \bm{j}^{\uparrow} \rangle - \langle \bm{j}^{\downarrow} \rangle$, where $\langle \bm{j}^{\sigma} \rangle$ ($\sigma = \uparrow, \downarrow$) is the spin-resolved electric current.
Furthermore, the local equilibrium current can be written with the spin-resolved orbital magnetization $\bm{M}^{\sigma}$ as $\langle \bm{j}_{\mathrm{s}}^{0, z} \rangle = \bm{\nabla} \times (\bm{M}^{\uparrow} - \bm{M}^{\downarrow})$.
In the presence of the gravitational potential, the local equilibrium current is written as
\begin{align}
\langle \bm{j}_{\mathrm{s}}^{\psi, z} \rangle
    & = (1 + \psi) \bm{\nabla} \times (\bm{M}^{\uparrow} - \bm{M}^{\downarrow})
\notag \\
    & = \bm{\nabla} \times \left\{ (\bm{M}^{\uparrow} - \bm{M}^{\downarrow}) (1 + \psi) \right\}
        - (\bm{\nabla} \psi) \times (\bm{M}^{\uparrow} - \bm{M}^{\downarrow})
.\label{eq:j^psi_con}
\end{align}
The last term in Eq.~(\ref{eq:j^psi_con}) contributes to the transport, as discussed in the thermal Hall effect~\cite{cooper1997,qin2011,shitade2014} and the spin Nernst effect for the Rashba model~\cite{dyrdal2016a,dyrdal2016}.
In the present paper, we investigate the generic cases where the spin is not necessarily conserved.

\subsection{\label{sec:sub:response_theory}Kubo formula for non-equilibrium spin current}
As shown before, the response coefficient consists of $\tilde{L}_{Q, ij}^{\alpha}$ and $\mathcal{J}_{i j}^{0, \alpha}$.
We first compute $\tilde{L}_{Q, ij}^{\alpha}$ using the Kubo formula.
Let us introduce the gravitational potential coupled with the local energy density as
\begin{align}
\mathcal{H}_{\mathrm{ext}} (t)
	& = Q (-\bq) \psi_{\bq} e^{-\zi \omega t}
,\end{align}
where $Q (\bq)$ and $\psi_{\bq}$ are the Fourier coefficients of the local energy density and the gravitational potential, respectively.
The response of the spin current is expressed as
\begin{align}
\langle j_{\mathrm{s},i}^{\alpha} (\bq, \omega) \rangle
	& = - \chi_{i 0}^{\alpha} (\bq, \omega) \psi_{\bq}
.\end{align}
According to the Kubo formula, $\chi_{i 0}^{\alpha} (\bq, \omega)$ is given by
\begin{align}
\chi_{i 0}^{\alpha} (\bq, \omega)
	& = \frac{\zi}{\hbar} \int_0^{\infty} \dd{t} e^{\zi (\omega + \zi 0) t} \langle [j_{\mathrm{s},i}^{\alpha} (\bq, t), Q (- \bq)] \rangle
,\end{align}
where $[ A, B ] = A B - B A$ is the commutator.
Integrating by parts and rearranging the forms, we obtain
\begin{align}
\chi_{i 0}^{\alpha} (\bq, \omega)
	& = \frac{\zi}{\hbar} \int_0^{\infty} \dd{t} \frac{e^{\zi (\omega + \zi 0) t} - 1}{\zi \omega} \langle [j_{\mathrm{s},i}^{\alpha} (\bq, t), \dot{Q} (- \bq)] \rangle
\notag
.\end{align}
We use the continuity equation for the local energy density
\begin{align}
\dot{Q} (\br) + \bm{\nabla} \cdot \bm{j}_{Q} (\br)
	& = 0
\label{eq:continuityEq}
,\end{align}
and then express $\chi_{i 0}^{\alpha} (\bq, \omega)$ as
\begin{align}
\chi_{i 0}^{\alpha} (\bq, \omega)
	& = \frac{\chi_{i j}^{\alpha} (\bq, \omega) - \chi_{i j}^{\alpha} (\bq, 0)}{\zi \omega} \zi q_j
\label{eq:resfunc}
.\end{align}
In Eq.~(\ref{eq:resfunc}), the response function is defined as
\begin{align}
\chi_{i j}^{\alpha} (\bq, \omega)
	& = \frac{\zi}{\hbar} \int_0^{\infty} \dd{t} e^{\zi (\omega + \zi 0) t} \langle [j_{\mathrm{s},i}^{\alpha} (\bq, t), j_{Q,j} (- \bq)] \rangle
.\end{align}
As a result, the response of the spin current is written as
\begin{align}
\langle j_{\mathrm{s},i}^{\alpha} (\bq, t) \rangle
	= \frac{\chi_{i j}^{\alpha} (\bq, \omega) - \chi_{i j}^{\alpha} (\bq, 0)}{\zi \omega} (- \zi q_j \psi_{\bq}) e^{- \zi \omega t}
.\end{align}
We consider the linear response to the uniform temperature gradients and thus take the limit of $\bq \to 0$ and then of $\omega \to 0$. We finally obtain
\begin{align}
\tilde{L}_{Q, ij}^{\alpha} 
	& = \lim_{\omega \to 0} \frac{\chi_{i j}^{\alpha} (\omega) - \chi_{i j}^{\alpha} (0)}{\zi \omega}
.\end{align}
where $\chi_{i j}^{\alpha} (\omega) = \chi_{i j}^{\alpha} (0, \omega)$.

\subsection{\label{sec:sub:evaluation}Evaluation of the non-equilibrium spin current}
The response function $\chi_{i j}^{\alpha} (\omega)$ is evaluated from the Matsubara correlation function defined as
\begin{align}
\tilde{\chi}_{i j}^{\alpha} (\zi \omega_{\lambda})
	& = \int_0^{\beta} \dd{\tau} e^{\zi \omega_{\lambda} \tau} \langle \mathrm{T}_{\tau} j_{\mathrm{s}, i}^{\alpha} (\tau) j_{Q, j} \rangle
\label{eq:chi_thermal}
,\end{align}
where $\omega_{\lambda} = 2 \pi \lambda \kB T$ is the Matsubara frequency of bosons, $j_{\mathrm{s}, i}^{\alpha} (\tau) = e^{\tau \tilde{\mathcal{H}}} j_{\mathrm{s}, i}^{\alpha} e^{- \tau \tilde{\mathcal{H}}}$ is the Heisenberg picture in the imaginary time $\tau$ with $\tilde{\mathcal{H}} = \int \dd{\br} (h (\br) - \mu n (\br))$, $\beta = 1 / \kB T$, $j_{\mathrm{s}, i}^{\alpha} = j_{\mathrm{s}, i}^{\alpha} (\bq = 0)$, and $j_{Q, j} = j_{Q, j} (\bq = 0)$.
$\chi_{i j}^{\alpha} (\omega)$ is associated with  $\tilde{\chi}_{i j}^{\alpha} (\zi \omega_{\lambda})$ by an analytical continuation in the complex frequency space as
\begin{align}
\chi_{i j}^{\alpha} (\omega)
	& = \tilde{\chi}_{i j}^{\alpha} (\hbar \omega + \zi 0)
.\end{align}

In the following, we assume a non-interacting electron system with the randomly distributed impurities described by
\begin{align}
\mathcal{H} = \int \dd{\br} h (\br) = H_0 + V_{\mathrm{imp}}
,\end{align}
where $H_0$ is a one-particle Hamiltonian with the translational symmetry and $V_{\mathrm{imp}}$ represents impurity potentials, respectively.
Because of the non-interacting electron system with the translational symmetry, the one-particle Hamiltonian can be diagonalized in the momentum space as $H_0 = \sum_{\bk} c^{\dagger}_{\bk} h_0 (\bk) c^{}_{\bk}$, where $c_{\bk}^{(\dagger)}$ is the annihilation (creation) operator of the electron with momentum $\bk$.
The impuriy potential has off-diagonal components in momentum space as $V_{\mathrm{imp}} = \sum_{\bk, \bq} \rho (\bq) u (\bq) c_{\bk+\bq}^{\dagger} c^{}_{\bk}$, where $\rho (\bq) = \sum_i e^{- \zi \bq \cdot \bm{R}_i}$ is the impurity density with the impurity positions $\bm{R}_i$ and the impurity potential strength $u (\bq)$.

The heat current density is expressed as
\begin{align}
j_{Q, i}
	& = \frac{1}{2} \sum_{\bk} \left( \dot{c}^{\dagger}_{\bk} v_i (\bk) c^{}_{\bk} - c^{\dagger}_{\bk} v_i (\bk) \dot{c}^{}_{\bk}\right)
\label{eq:jq}
,\end{align}
where $\dot{c}^{(\dag)}_{\bk}=d{c^{(\dag)}_{\bk}}/d\tau$ is the imaginary-time derevative and $v_i (\bk) = (1/\hbar) (\partial h_0 (\bk) / \partial k_i)$ is the velocity operator in the Fourier space.
This expression is obtained from the Heisenberg equation of the local energy density combined with the continuity equation of Eq.~(\ref{eq:continuityEq}).
Jonson and Mahan~\cite{jonson1980} showed that this type of equation (\ref{eq:jq}) holds for the impurity potential and for part of the electron-phonon interaction to prove the Mott formula for the Seebeck coefficient.
This method was extended to the Hubbard interaction by Kontani~\cite{kontani2003} and for more general cases by one of the authors~\cite{ogata2019} to clarify the range of validity of Mott's formula.
For the derivation in the three-dimensional Dirac electron system, see Appendix~A in Ref.~\cite{fujimoto2022a}.

The spin current density is expressed as
\begin{align}
j_{\mathrm{s}, i}^{\alpha}
	& = \frac{\hbar}{2} \sum_{\bk} c^{\dagger}_{\bk} v_{\mathrm{s}, i}^{\alpha} (\bk) c^{}_{\bk}
\label{def:js}
,\end{align}
where $v_{\mathrm{s}, i}^{\alpha} (\bk)$ is the velocity of the spin current.
Note that we do not assume the explicit form of $v_{\mathrm{s}, i}^{\alpha} (\bk)$.
We also note that one can discuss other responses to the temperature gradient by replacing $(\hbar/2) v_{\mathrm{s}, i}^{\alpha} (\bk)$ in Eq.~(\ref{def:js}) with the focused quantities, such as the spin density $(\hbar/2) \sigma^{\alpha}$ and the electric current $- e v_i (\bk)$.

Substituting Eq.~(\ref{eq:jq}) and (\ref{def:js}) into Eq.~(\ref{eq:chi_thermal}) and then using Wick's theorem, we have
\begin{widetext}
\begin{align}
\tilde{\chi}_{i j}^{\alpha} (\zi \omega_{\lambda})
	& = + \frac{\hbar}{4} \int_0^{\beta} \dd{\tau} e^{\zi \omega_{\lambda} \tau}
	 \sum_{\bk, \bk'} \tr \left[
			v_{\mathrm{s}, i}^{\alpha} (\bk) \langle \mathrm{T}_{\tau} c^{}_{\bk} (\tau) \dot{c}_{\bk'}^{\dagger} \rangle
			v_j (\bk') \mathcal{G}_{\bk', \bk} (- \tau)
			- v_{\mathrm{s}, i}^{\alpha} (\bk) \mathcal{G}_{\bk, \bk'} (\tau)
			v_j (\bk') \langle \mathrm{T}_{\tau} \dot{c}^{}_{\bk'} c_{\bk}^{\dagger} (\tau) \rangle
		\right]
\label{eq:chiM}
,\end{align}
\end{widetext}
where $\mathcal{G}_{\bk, \bk'} (\tau - \tau') = - \langle \mathrm{T}_{\tau} c^{}_{\bk} (\tau) c^{\dagger}_{\bk'} (\tau') \rangle_{\mathcal{H}} $ is the imaginary-time Green function of $\mathcal{H} = H_0 + V_{\mathrm{imp}}$ and $\mathrm{tr}$ represents the trace over spin degrees of freedom.
Note that 
$\mathcal{G}_{\bk, \bk'} (\tau)$ is not a diagonal in momentum space due to $V_{\mathrm{imp}}$.
We use
\begin{subequations}
\begin{align}
\langle \mathrm{T}_{\tau} c^{}_{\bk} (\tau) \dot{c}_{\bk'}^{\dagger} \rangle
	& = \frac{d}{d \tau} \mathcal{G}_{\bk, \bk'} (\tau) + \delta_{\bk \bk'} \delta (\tau)
, \\
\langle \mathrm{T}_{\tau} \dot{c}^{}_{\bk} c_{\bk'}^{\dagger} (\tau) \rangle
	& = \frac{d}{d \tau} \mathcal{G}_{\bk, \bk'} (- \tau) - \delta_{\bk \bk'} \delta (\tau)
,\end{align}
\end{subequations}
to rewrite Eq.~(\ref{eq:chiM}) as 
\begin{align}
\tilde{\chi}_{i j}^{\alpha} (\zi \omega_{\lambda})
	& = - \frac{\hbar}{2 \beta V^2} \sum_n \sum_{\bk, \bk'} \tr \left[
		v_{\mathrm{s}, i}^{\alpha} (\bk) \mathcal{G}_{\bk, \bk'} (\zi \epsilon_n^{+})
\notag \right. \\ & \hspace{3em} \left. \times 
		\left( \frac{\zi \epsilon_n^{+} + \zi \epsilon_n}{2} \right) v_j (\bk') \mathcal{G}_{\bk', \bk} (\zi \epsilon_n)
	\right]
 \label{eq:chiM1}
,\end{align}
where $\epsilon_n = (2 n + 1) \pi \kB T$ is the Matsubara frequency of fermions and $\epsilon_n^{+} = \epsilon_n + \omega_{\lambda}$.
In Eq.~(\ref{eq:chiM1}), the terms proportional to $\delta (\tau)$ are neglected because they do not depend on $\zi \omega_{\lambda}$ and thus independent of $\omega$ after taking the analytic continuation.

Now, we take the statistical average on the impurity positions defined as
\begin{align}
\langle A (\{ \bm{R}_i \}) \rangle_{\mathrm{av}}
	& = \prod_{i} \int \frac{\dd{\bm{R}_i}}{V} A (\{ \bm{R}_i \})
,\end{align}
where $\{ \bm{R}_i \}$ is the set of the positions of the impurity potentials.
With this statistical average, we define the impurity averaged Matsubara Green function as
\begin{align}
G_{\bk} (\zi \epsilon_n) = \langle \mathcal{G}_{\bk, \bk} (\zi \epsilon_n) \rangle_{\mathrm{av}} = \{ \zi \epsilon_n - h_0 (\bk) - \Sigma_{\bk} (\zi \epsilon_n) \}^{-1}
,\end{align}
with the self energy due to the impurity potential.
Using 
$G_{\bk} (\zi \epsilon_n)$, we express the correlation function as
\begin{align}
\langle \tilde{\chi}_{i j}^{\alpha} (\zi \omega_{\lambda}) \rangle_{\mathrm{av}}
	& = -\frac{\hbar}{2 \beta V} \sum_n \sum_{\bk} \tr \left[
		\Lambda_{\mathrm{s}, i}^{\alpha} (\bk; \zi \epsilon_n, \zi \epsilon_n^{+}) G_{\bk} (\zi \epsilon_n^{+})
\notag \right. \\ & \hspace{3em} \left. \times 
		\left( \zi \epsilon_n + \frac{\zi \omega_{\lambda}}{2} \right) v_j (\bk) G_{\bk} (\zi \epsilon_n)
	\right]
.\end{align}
where $\Lambda_{\mathrm{s}, i}^{\alpha} (\bk; \zi \epsilon_n, \zi \epsilon_n^{+})$ is the full vertex of the spin currnet including the vertex correction corresponding to the impurity self energy.

Nest, we rewrite the Matsubara summation using the contour integral and then take the analytic continuation $\zi \omega_{\lambda} \to \hbar \omega + \zi 0$.
The resultant expression is
\begin{align}
\langle \chi_{i j}^{\alpha} (\omega) \rangle_{\mathrm{av}}
	& = \frac{\hbar}{2} \int_{-\infty}^{\infty} \frac{\dd{\epsilon}}{2 \pi \zi} \left[
		\bigl( f (\epsilon_{+}) - f (\epsilon_{-}) \bigr) \epsilon \varphi_{i j}^{\R \A, \alpha} (\epsilon_{+}, \epsilon_{-})
\right. \notag \\ & \hspace{1em} \left.
		+ f (\epsilon_{-}) \epsilon \varphi_{i j}^{\R\R, \alpha} (\epsilon_{+}, \epsilon_{-})
		- f (\epsilon_{+}) \epsilon \varphi_{i j}^{\A\A, \alpha} (\epsilon_{+}, \epsilon_{-})
	\right]
  \label{eq:chi1}
,\end{align}
with $f (\epsilon) = (e^{\beta \epsilon} + 1)^{-1}$ and $\epsilon_{\pm} = \epsilon \pm \hbar \omega / 2$.
In Eq.~(\ref{eq:chi1}),
\begin{align}
\varphi_{i j}^{\X\Y, \alpha} (\epsilon, \epsilon')
	& = \frac{1}{V} \sum_{\bk} \tr \Biggl[
		\Lambda_{\mathrm{s}, i}^{\Y \X, \alpha} (\bk; \epsilon', \epsilon) G^{\X}_{\bk} (\epsilon)
		v_j (\bk) G^{\Y}_{\bk} (\epsilon')
	\Biggr]
,\end{align}
is defined, where $\X, \Y \in \{ \R, \A \}$ and $G^{\R (\A)}_{\bk} (\epsilon)$ is the retarded (advanced) Green function.
Here, $\Lambda_{\mathrm{s}, i}^{\A \R, \alpha} (\bk; \epsilon', \epsilon)$ is the full vertex of the spin current obtained by taking the analytic continuations, $\zi \epsilon'_n \to \epsilon' - \zi 0$ and $\zi \epsilon_n \to \epsilon + \zi 0$ in $\Lambda_{\mathrm{s}, i}^{\alpha} (\bk; \zi \epsilon'_n, \zi \epsilon_n)$.
Similarly, $\Lambda_{\mathrm{s}, i}^{\R \R, \alpha} (\bk; \epsilon', \epsilon)$ ($\Lambda_{\mathrm{s}, i}^{\A \A, \alpha} (\bk; \epsilon', \epsilon)$) is obtained by taking the analytic continuations, $\zi \epsilon'_n \to \epsilon' + \zi 0$ and $\zi \epsilon_n \to \epsilon + \zi 0$ ($\zi \epsilon'_n \to \epsilon' - \zi 0$ and $\zi \epsilon_n \to \epsilon - \zi 0$) in $\Lambda_{\mathrm{s}, i}^{\alpha} (\bk; \zi \epsilon'_n, \zi \epsilon_n)$.
By expanding for $\omega$ in $\chi_{i j}^{\alpha} (\omega)$, we finally obtain
\begin{align}
\tilde{L}_{i j}^{\alpha}
	& = \frac{\hbar^2}{4 \pi} \int_{- \infty}^{\infty} \dd{\epsilon} \biggl[
		\left(- \frac{\partial f}{\partial \epsilon} \right) \epsilon F_{1, ij}^{\alpha} (\epsilon + \mu)
		+ f (\epsilon) \epsilon F_{2, ij}^{\alpha} (\epsilon + \mu)
	\biggr]
\label{eq:KuboLuttingerFinalL}
,\end{align}
where
\begin{align}
F_{1, ij}^{\alpha} (\epsilon + \mu)
    & = \varphi_{i j}^{\R \A, \alpha} (\epsilon, \epsilon)
        - \frac{1}{2} \left\{
            \varphi_{i j}^{\R \R, \alpha} (\epsilon, \epsilon)
            + \varphi_{i j}^{\A \A, \alpha} (\epsilon, \epsilon)
        \right\}
\label{eq:F1}
, \\
F_{2, ij}^{\alpha} (\epsilon + \mu)
    & = - \frac{1}{2} (\partial_{\epsilon} - \partial_{\epsilon'}) \left\{
            \varphi_{i j}^{\R \R, \alpha} (\epsilon, \epsilon')
            - \varphi_{i j}^{\A \A, \alpha} (\epsilon, \epsilon')
	   \right\} \bigg|_{\epsilon' \to \epsilon}
\label{eq:F2}
\end{align}

Let us show that the non-equilibrium current diverges at $T= 0$.
At low temperatures, we expand
\begin{align}
\int_{-\infty}^{\infty} \dd{\epsilon} \left( - \frac{\partial f (\epsilon)}{\partial \epsilon} \right) \epsilon F_{1, ij}^{\alpha} (\epsilon + \mu)
	& \simeq \frac{\pi^2 (\kB T)^2}{3} \left. \frac{\partial F_{1, ij}^{\alpha} (\epsilon)}{\partial \epsilon} \right|_{\epsilon = \eF}
, \\
\int_{-\infty}^{\infty} \dd{\epsilon} f (\epsilon) \epsilon F_{2, ij}^{\alpha} (\epsilon + \mu)
	& \simeq \int_{-\infty}^{\eF} \dd{\epsilon} (\epsilon - \eF) F_{2, ij}^{\alpha} (\epsilon)
\end{align}
with the Fermi energy $\eF$ defined as $\eF=\mu(T=0)$.
Hence, we obtain the expression of $\tilde{L}_{i j}^{\alpha}$ at low temperatures as
\begin{align}
\tilde{L}_{i j}^{\alpha}
	& = \frac{\hbar^2}{4 \pi} \left\{
		\frac{\pi^2 (\kB T)^2}{3} \left. \frac{\partial F_{1, ij}^{\alpha} (\epsilon)}{\partial \epsilon} \right|_{\epsilon = \eF}
\right. \notag \\ & \hspace{6em} \left.
        + \int_{-\infty}^{\eF} \dd{\epsilon} (\epsilon - \eF) F_{2, ij}^{\alpha} (\epsilon)
	\right\}
\label{eq:tilde_L_lowT}
.\end{align}
We note that the second term does not depend on $T$ at low temperatures so that $\tilde{L}_{i j}^{\alpha} / T$ diverges as $T \to 0$.
Since thermodynamics requires that the response to temperature gradients should vanish at $T = 0$, the above standard procedures of the Kubo formula for the thermal response seemingly lead to the unphysical result.
As shown in Sec.~\ref{sec:sub:Einsteins_relation_and_Luttingers_method}, we need to consider the contribution from the local equilibrium current, $\mathcal{J}_{i j}^{0, \alpha}$, which precisely removes the unphysical divergence as shown below.

To see this more precisely, we show the expression of $\tilde{L}_{i j}^{\alpha} $ in the clean limit, where no impurity potential is present.
In this limit, the Green function can be expressed using the Bloch eigenstate $\ket{n \bk}$ with the band index $n$ and the crystal momentum vector $\bk$ as
\begin{align}
G^{\R/\A}_{\bk} (\epsilon)
    & = \sum_n \frac{ \ket{n \bk} \bra{n \bk} }{\epsilon + \mu - \epsilon_{n \bk} \pm \zi 0}
\label{eq:expansion_G}
,\end{align}
where $\epsilon_{n \bk}$ is the corresponding eigenvalue.
In this limit, we can compute the above calculation exactly (see Appendix~\ref{apx:cleanlimit} for the calculation detail) and obtain
\begin{align}
\tilde{L}_{i j}^{\alpha}
    & = - \frac{1}{V} \sum_{n, \bk} f_{\mathrm{FD}} (\epsilon_{n \bk}) \left( (\epsilon_{n \bk} - \mu) \Omega_{n \bk, i j}^{\alpha} - \frac{1}{2} m_{n \bk, i j}^{\alpha} \right)
\label{eq:clean_limit_L_tilde}
,\end{align}
where $f_{\mathrm{FD}} (\epsilon) = \{ e^{(\epsilon - \mu) / \kB T} + 1 \}^{-1}$ is the Fermi-Dirac distribution function, and $\Omega_{n \bk, i j}^{\alpha}$ is the Berry curvature-like quantity (sometimes called the spin Berry curvature~\cite{taguchi2020,lau2023}),
\begin{align}
\Omega_{n \bk, i j}^{\alpha}
    & = \frac{\zi \hbar^2}{2} \sum_{m \neq n} \frac{\braket{n | v_{\mathrm{s}, i}^{\alpha} | m} \braket{m | v_j | n} - \braket{n | v_j | m} \braket{m | v_{\mathrm{s}, i}^{\alpha} | n}}{(\epsilon_{n \bk} - \epsilon_{m \bk})^2}
\label{eq:Omega_nk}
\end{align}
and $ m_{n \bk, i j}^{\alpha}$ is the spin-dependent magnetic moment,
\begin{align}
m_{n \bk, i j}^{\alpha}
    & = \frac{\zi \hbar^2}{2} \sum_{m \neq n} \frac{\braket{n | v_{\mathrm{s}, i}^{\alpha} | m} \braket{m | v_j | n} - \braket{n | v_j | m} \braket{m | v_{\mathrm{s}, i}^{\alpha} | n}}{\epsilon_{n \bk} - \epsilon_{m \bk}}
\label{eq:m_nk}
.\end{align}
When the spin is conserved, the Berry curvature-like quantity and the spin-dependent magnetic moment are expressed as
\begin{align}
\Omega_{n \bk, i j}^{z}
    & = \epsilon^{i j l} \frac{b_{n \bk, l}^{\uparrow} - b_{n \bk, l}^{\downarrow}}{2}
, \\
m_{n \bk, i j}^{z}
    & = \epsilon^{i j l} \frac{m_{n \bk, l}^{\uparrow} - m_{n \bk, l}^{\downarrow}}{2}
,\end{align}
where $\bm{b}_{n \bk}^{\sigma}$ and $\bm{m}_{n \bk}^{\sigma}$ with $\sigma = \uparrow, \downarrow$ are the spin-resolved Berry curvature and magnetic moment, respectively.

\subsection{\label{sec:sub:local_equilibrium}Local equilibrium spin current}
Next, we evaluate the local equilibrium current $\langle j_{\mathrm{s}, i}^{0, \alpha} \rangle = \nabla_j \mathcal{J}_{i j}^{0, \alpha} [\mu (\br), T (\br)]$.
We hereafter write $\langle j_{\mathrm{s}, i}^{0, \alpha} \rangle$ as $\langle j_{\mathrm{s}, i}^{\alpha} \rangle_{\mathrm{leq}}$ to emphasize the local equilibrium quantity.
To obtain the functional $\mathcal{J}_{i j}^{0, \alpha} [\mu (\br), T (\br)]$, we consider a situation in which the chemical potential $\mu (\br)$ and the temperature $T (\br)$ have the form
\begin{align}
\mu (\br)
	= \mu_0 + \delta \mu (\br)
, & \qquad
1 / T (\br)
	= 1 / T_0 + \delta ( 1 / T (\br) )
.\end{align}
Here, $\mu_0$ and $T_0$ are the chemical potential and temperature in the global equilibrium.
Then, the deviation of the local equilibrium current $\langle j_{\mathrm{s}, i}^{0, \alpha} \rangle$ 
is calculated from the static response theory~\cite{kubo1991} as 
\begin{align}
\delta\langle j_{\mathrm{s}, i}^{\alpha} (\br) \rangle_{\mathrm{leq}}
	& = 
		 \int \dd{\br'} \phi_{c, i}^{\alpha} (\br, \br') \delta \mu (\br')
\notag \\ & \hspace{1em}
		- \int \dd{\br'} \phi_{Q, i}^{\alpha} (\br, \br') T_0 \delta \left( \frac{1}{T (\br')} \right)
,\end{align}
with
\begin{subequations}
\begin{align}
\phi_{c, i}^{\alpha} (\br, \br')
	& = \int_0^{\beta_0} \dd{\tau} \langle \varDelta j_{\mathrm{s},i}^{\alpha} (\br, \tau) \varDelta n (\br') \rangle_0
, \\
\phi_{Q, i}^{\alpha} (\br, \br')
	& = \int_0^{\beta_0} \dd{\tau} \langle \varDelta j_{\mathrm{s},i}^{\alpha} (\br, \tau) [\varDelta h (\br') - \mu_0 \varDelta n(\br ')] \rangle_0
,\end{align}
\label{eqs:static_response_function}
\end{subequations}
where $\langle \,\cdots \rangle_{0}$ is the expectation value in global equilibrium, $\beta_0 = 1 / \kB T_0$, $\varDelta j_{\mathrm{s}, i}^{\alpha}(\br) = j_{\mathrm{s}, i}^{\alpha}(\br) - \langle j_{\mathrm{s}, i}^{\alpha} (\br) \rangle_0$, $\varDelta n (\br) = n(\br) - \langle n(\br) \rangle_0$, and $\varDelta h (\br) = h (\br) - \langle h (\br) \rangle_0$.
Note the effect of $\delta \mu (\br)$ appears through $-\mu(\br) n(\br)$ in the static expectation values and that of $\delta(1/ T(\br))$ through $\beta(\br) [h(\br) - \mu n(\br)]$, which leads to Eq.~(\ref{eqs:static_response_function}). 

Assuming the uniform system, we perform the Fourier transformation to obtain
\begin{align}
\delta\langle j_{\mathrm{s}, i}^{\alpha} (\br) \rangle_{\mathrm{leq}}
	& = 
		 \frac{1}{V} \sum_{\bq} e^{\zi \bq \cdot \br} \phi_{c, i}^{\alpha} (\bq) \delta \mu_{\bq}
  \notag \\ & \hspace{1em}
		- \frac{1}{V} \sum_{\bq} e^{\zi \bq \cdot \br} \phi_{Q, i}^{\alpha} (\bq) T_0 \delta \left( \frac{1}{T} \right)_{\bq}
.\end{align}
In the similar way to calculate $\chi_{ij}^{\alpha}(\omega)$ in the Kubo formula, we can show
\begin{widetext}
\begin{align}
\phi_{c, i}^{\alpha}(\bq)
	& = \int_0^{\beta_0} \dd{\tau} \langle \varDelta j_{\mathrm{s},i}^{\alpha} (\bq, \tau) \varDelta n (-\bq) \rangle_0 \nonumber
 \\
	& = \frac{\hbar}{2V} \int_{-\infty}^{\infty} \frac{\dd{\epsilon}}{2 \pi \zi} f (\epsilon) \sum_{\bk} \tr \biggl[
			\Lambda_{\mathrm{s},i}^{\R\R, \alpha}(\bk,\bq, \epsilon,\epsilon) 
        G^{\R}_{\bk+\bq/2}(\epsilon) 
        G^{\R}_{\bk-\bq/2}(\epsilon) 
        -\Lambda_{\mathrm{s},i}^{\A\A, \alpha}(\bk,\bq, \epsilon,\epsilon) 
        G^{\A}_{\bk+\bq/2}(\epsilon) 
        G^{\A}_{\bk-\bq/2}(\epsilon) \biggr]
\label{eq:phi_cFull}
.\end{align}
\end{widetext}
By using the following relation
\begin{align}
h(\bq, \tau)-\mu n(\bq, \tau)
    & = \frac{1}{2} \left( \partial_{\tau}-\partial_{\tau'} \right)
        \sum_{\bk} c_{\bk-\frac{\bq}{2}}^{\dagger} (\tau) c_{\bk+\frac{\bq}{2}} (\tau')
        \biggl|_{\tau' \rightarrow \tau}
,\end{align}
we can see that $\phi_{Q, i}^{\alpha}(\bq)$ is obtained by replacing $f(\epsilon)$ in Eq.~(\ref{eq:phi_cFull}) with $f(\epsilon) \epsilon$.

Then, expanding for small $\bq$ as
\begin{align*}
\phi_{c, i}^{\alpha} (\bq)
	\simeq \phi_{c, i}^{\alpha} (0)
		+ \zi q^j \phi_{c, i j}^{\alpha}
, 
\\
\phi_{Q, i}^{\alpha} (\bq)
	\simeq \phi_{Q, i}^{\alpha} (0)
		+ \zi q^j \phi_{Q, i j}^{\alpha}
,\end{align*}
we can see that $\phi_{c, i}^{\alpha} (0)=\phi_{Q, i}^{\alpha} (0)=0$ since the $\bq = 0$ components are nothing but the quantities in the global equilibrium.
In the linear order of $\bq$, considering that the vertex function $\Lambda_{\mathrm{s},i}^{XX, \alpha}$ contains $G^{\X}_{\bk+\bq/2}(\epsilon)$ and $G^{\X}_{\bk-\bq/2}(\epsilon)$, we obtain
\begin{align}
\phi_{c, i j}^{\alpha}
	& = \int_{-\infty}^{\infty} \frac{\dd{\epsilon}}{2 \pi \zi} f (\epsilon) \Phi_{ij}^{\alpha} (\epsilon + \mu_0)
\label{eq:phi_c}
,\end{align}
\begin{align}
\phi_{Q, i j}^{\alpha}
	& = \int_{-\infty}^{\infty} \frac{\dd{\epsilon}}{2 \pi \zi} f (\epsilon) \epsilon \Phi_{ij}^{\alpha} (\epsilon + \mu_0)
\label{eq:phi_Q}
,\end{align}
with $f (\epsilon) = (e^{\beta_0 \epsilon} + 1)^{-1}$.
In Eqs.~(\ref{eq:phi_c}) and~(\ref{eq:phi_Q}), $\Phi_{ij}^{\alpha} (\epsilon + \mu_0)$ is defined as
\begin{widetext}
\begin{align}
\Phi_{ij}^{\alpha} (\epsilon + \mu_0)
	& = \frac{\hbar}{4 \zi V} \sum_{\bk} \tr\left[
			\Lambda_{\mathrm{s},i}^{\R\R, \alpha} \left( \partial_j G^{\R}_{\bk} \right) \Lambda_0^{\R\R} G^{\R}_{\bk}
			- \Lambda_{\mathrm{s},i}^{\R\R,\alpha} G^{\R}_{\bk} \Lambda_0^{\R\R} \left( \partial_j G^{\R}_{\bk} \right)
\right. \notag \\ & \hspace{5em} \left.
			- \Lambda_{\mathrm{s},i}^{\A\A, \alpha} \left( \partial_j G^{\A}_{\bk} \right) \Lambda_0^{\A\A} G^{\A}_{\bk}
			+ \Lambda_{\mathrm{s},i}^{\A\A, \alpha} G^{\A}_{\bk} \Lambda_0^{\A\A} \left( \partial_j G^{\A}_{\bk} \right)
		\right]
,\end{align}
\end{widetext}
with $\partial_j G^{\R}_{\bk} = \partial G^{\R}_{\bk} / \partial k^j$ and $\Lambda_{\mathrm{s},i}^{\R\R, \alpha} = \Lambda_{\mathrm{s},i}^{\R\R, \alpha} (\bk; \epsilon, \epsilon)$. 
The vertex $\Lambda_0^{\R\R}$ is the full vertex of the electron number and $\Lambda_0^{\A\A}$ is obtained by replacing $G^{\R}_{\bk}$ with $G^{\A}_{\bk}$ in $\Lambda_0^{\R\R}$.

Using the Ward-Takahashi identities, 
\begin{align}
G^{\R}_{\bk} \Lambda_0^{\R\R} G^{\R}_{\bk}
	= - \partial_{\epsilon} G^{\R}_{\bk}
, & \qquad
\hbar G^{\R}_{\bk} \Lambda_{v, i}^{\R\R} G^{\R}_{\bk}
	= \partial_{i} G^{\R}_{\bk}
, \\
G^{\A}_{\bk} \Lambda_0^{\A\A} G^{\A}_{\bk}
	= - \partial_{\epsilon} G^{\A}_{\bk}
, & \qquad
\hbar G^{\A}_{\bk} \Lambda_{v, i}^{\A\A} G^{\A}_{\bk}
	= \partial_{i} G^{\A}_{\bk}
,\end{align}
we find
\begin{align}
\tr
	& \left[ \Lambda_{\mathrm{s},i}^{\R\R, \alpha} \left( \partial_j G^{\R}_{\bk} \right) \Lambda_0^{\R\R} G^{\R}_{\bk}
		- \Lambda_{\mathrm{s},i}^{\R\R, \alpha} G^{\R}_{\bk} \Lambda_0^{\R\R} \left( \partial_j G^{\R}_{\bk} \right)
\right]
\notag \\
	& = - \hbar \, \tr \left[
		\Lambda_{\mathrm{s},i}^{\R\R, \alpha} G^{\R}_{\bk} \Lambda_{v, j}^{\R\R} \left( \partial_{\epsilon} G^{\R}_{\bk} \right)
		- \Lambda_{\mathrm{s},i}^{\R\R, \alpha} \left( \partial_{\epsilon} G^{\R}_{\bk} \right) \Lambda_{v, j}^{\R\R} G^{\R}_{\bk}
	\right]
,\end{align}
where $\Lambda_{v, i}^{\X\X}$ ($\X \in {\R, \A}$) is the full velocity vertex.
A similar form is obtained for the advanced part by replacing $\R \to \A$.
Hence, we obtain
\begin{widetext}
\begin{align}
\Phi_{ij}^{\alpha} (\epsilon + \mu_0)
	& = - \frac{\hbar^2}{4 \zi V} \sum_{\bk} \tr \left[
		\Lambda_{\mathrm{s},i}^{\R\R, \alpha} G^{\R}_{\bk} \Lambda_{v, j}^{\R\R} \left( \partial_{\epsilon} G^{\R}_{\bk} \right)
		- \Lambda_{\mathrm{s},i}^{\R\R, \alpha} \left( \partial_{\epsilon} G^{\R}_{\bk} \right) \Lambda_{v, j}^{\R\R} G^{\R}_{\bk}
		- \Lambda_{\mathrm{s},i}^{\A\A, \alpha} G^{\A}_{\bk} \Lambda_{v, j}^{\A\A} \left( \partial_{\epsilon} G^{\A}_{\bk} \right)
		+ \Lambda_{\mathrm{s},i}^{\A\A, \alpha} \left( \partial_{\epsilon} G^{\A}_{\bk} \right) \Lambda_{v, j}^{\A\A} G^{\A}_{\bk}
	\right]
\label{eq:Phi_ij}
.\end{align}
Decomposing the full velocity vertex $\Lambda_{v, j}^{\R\R}$ and $\Lambda_{v, j}^{\A\A}$ and then recollecting the terms, we have
\begin{align}
\Phi_{ij}^{\alpha} (\epsilon + \mu_0)
	& = \frac{\hbar^2}{4 \zi V} (\partial_{\epsilon} - \partial_{\epsilon'})
    \sum_{\bk} \tr \left[
		\Lambda_{\mathrm{s}, i}^{\R\R, \alpha} (\bk; \epsilon, \epsilon') G^{\R}_{\bk} (\epsilon)
		v_j (\bk) G^{\R}_{\bk} (\epsilon')
		- \Lambda_{\mathrm{s}, i}^{\A\A, \alpha} (\bk; \epsilon, \epsilon') G^{\A}_{\bk} (\epsilon)
		v_j (\bk) G^{\A}_{\bk} (\epsilon')
	\right] \Big|_{\epsilon' \to \epsilon}
\notag \\
	& = \frac{\zi \hbar^2}{2} F_{2, ij}^{\alpha} (\epsilon+\mu_0)
.\end{align}
\end{widetext}
Hence, we obtain
\begin{align}
\delta \langle j_{\mathrm{s}, i}^{\alpha} (\br) \rangle_{\mathrm{leq}}
	& = 
	     \frac{\zi\hbar^2}{2\beta_0} \int_{-\infty}^{\infty} \frac{\dd{\epsilon}}{2 \pi \zi} f (\epsilon) F_{2, ij}^{\alpha} (\epsilon+\mu_0)
\notag \\ & \hspace{2em} \times
		\nabla_j \left\{
			\frac{\delta \mu (\br)}{\kB T_0}
			- \epsilon \delta \left( \frac{1}{\kB T (\br)} \right)
		\right\}
.\end{align}
By changing the variable as $\epsilon+\mu_0 \to \epsilon$ and writing $\{ \cdots \}$ as $- \delta ( (\epsilon - \mu (\br)) / \kB T (\br) )$, we find
\begin{align}
\beta_0 \delta \langle j_{\mathrm{s}, i}^{\alpha} (\br) \rangle_{\mathrm{leq}}
	& = \nabla_j \frac{\zi\hbar^2}{2} \int_{-\infty}^{\infty} \frac{\dd{\epsilon}}{2 \pi \zi} F_{2,ij}^{\alpha} (\epsilon)
\notag \\ & \hspace{2em} \times
	\delta \left( \ln \left[
		1 + \exp \left( - \frac{\epsilon - \mu (\br)}{\kB T (\br)} \right)
	\right] \right)
\label{eq:DeviationFinal}
,\end{align}
where we have assumed that the $\mu$- and $T$-dependences of $F_{2,ij}^{\alpha} (\epsilon)$ is negligible~\footnote{For the non-interacting systems, $F_{2, ij}^{\alpha}$ does not contain $\mu$ nor $T$, but it can depend on $\mu$ and $T$ in general interacting cases.}.
Therefore, we obtain
\begin{align}
\langle j_{\mathrm{s}, i}^{\alpha} (\br) \rangle_{\mathrm{leq}}
	& = \nabla^j \mathcal{J}_{ij}^{0, \alpha} [\mu (\br), T (\br)]
,\end{align}
where
\begin{align}
\mathcal{J}_{ij}^{0, \alpha} [\mu, T]
	& = \frac{\hbar^2 \kB T}{4\pi} \int_{-\infty}^{\infty} \dd{\epsilon} F_{2,ij}^{\alpha} (\epsilon)
		\ln \left[
			1 + \exp \left( - \frac{\epsilon - \mu}{\kB T} \right)
		\right]
\label{eq:J_ij}
.\end{align}

At $T=0$, this expression becomes
\begin{align}
\mathcal{J}_{ij}^{0, \alpha} [\eF, 0]
	& = - \frac{\hbar^2}{4\pi} \int_{-\infty}^{\eF} \dd{\epsilon} (\epsilon-\eF) F_{2,ij}^{\alpha} (\epsilon)
.\end{align}
As mentioned before, this contribution from the local equilibrium current exactly cancels with $T=0$ limit of $\tilde L_{ij}^{\alpha}$, i.e., this is significantly important to remove the unphysical divergence of the spin Nernst current at $T=0$.

For finite temperatures, we use a relation
\begin{align}
& \kB T \ln \left[ 1 + \exp \left( - \frac{\epsilon - \mu}{\kB T} \right) \right]
\notag \\ & \hspace{1em}
    = \int^{\infty}_{\epsilon-\mu} \dd{\epsilon'} f(\epsilon')
\notag \\ & \hspace{1em}
    = - (\epsilon - \mu) f(\epsilon - \mu)
        - \int^{\infty}_{\epsilon-\mu} \dd{\epsilon'} \epsilon' \frac{\partial f}{\partial \epsilon'}
\label{eq:NiceTrick}
.\end{align}
Substituting this relation into in Eq.~(\ref{eq:J_ij}) and then changing the variable $\epsilon - \mu' \to \epsilon$ in the $\epsilon$ integral. We obtain
\begin{align}
& \mathcal{J}_{ij}^{0, \alpha} [\mu, T]
\notag \\ & \hspace{1em}
    = -\frac{\hbar^2}{4\pi} \int_{-\infty}^{\infty} \dd{\epsilon} F_{2,ij}^{\alpha} (\epsilon+\mu)
    \left[ f(\epsilon)\epsilon + \int_{\epsilon}^{\infty} \dd{\epsilon'} \epsilon' \frac{\partial f}{\partial \epsilon'}\right]
\notag \\ & \hspace{1em}
    = -\frac{\hbar^2}{4\pi} \int_{-\infty}^{\infty} \dd{\epsilon} 
    \left[ f(\epsilon)\epsilon F_{2,ij}^{\alpha} (\epsilon+\mu) + \frac{\partial f}{\partial \epsilon} \epsilon \mathcal{F}_{2, ij}^{\alpha} (\epsilon + \mu) \right] 
\label{eq:EquivFinalL}
.\end{align}
where $\mathcal{F}_{2, ij}^{\alpha} (\epsilon + \mu)$ is defined as
\begin{align}
\mathcal{F}_{2, ij}^{\alpha} (\epsilon + \mu)
    & \equiv \int_{-\infty}^{\epsilon} \dd{\epsilon'} F_{2, ij}^{\alpha} (\epsilon' + \mu)
,\end{align}
and the chage of the integral $\int_{-\infty}^\infty \dd{\epsilon} \int_{\epsilon}^\infty \dd{\epsilon'} = \int_{-\infty}^\infty \dd{\epsilon'} \int^{\epsilon'}_{-\infty} \dd{\epsilon}$ has been used in the second term.
The first term exactly cancels with the second term of the nonequilibrium spin current obtained by the Kubo formula in Eq.~(\ref{eq:KuboLuttingerFinalL}).
This indicates that the local equilibrium current successfully removes the divergence of the non-equilibrium spin current at low temperatures.

Finally, total of Eqs.~(\ref{eq:KuboLuttingerFinalL}) and (\ref{eq:EquivFinalL}) leads to the following expression of the response coefficient
\begin{align}
L_{Q, i j}^{\alpha}
	& = \frac{\hbar^2}{4 \pi} \int_{-\infty}^{\infty} \dd{\epsilon}
		\left( - \frac{\partial f_{\mathrm{FD}}}{\partial \epsilon} \right) (\epsilon - \mu)
		\left[
			F_{1, ij}^{\alpha} (\epsilon)
			+ \mathcal{F}_{2, ij}^{\alpha} (\epsilon)
	\right]
\label{eq:precise_LQ}
.\end{align}
Note that our discussion above only relies on the Ward-Takahashi identities and hence is generic.
The Ward-Takahashi identities guarantee that the computed spin Nernst current does not involve the unphysical divergence at $T=0$ as long as the calculation does not break the conservation law.

\subsection{\label{sec:clean_limit_and_streda_formula}Clean limit and St\v{r}eda formula}
Here, we examine the results obtained in the preceding sections in the clean limit.
With the Green function represented by the Bloch eigenstate [Eq.~(\ref{eq:expansion_G})], the local equilibrium current in the clean system is expressed as
\begin{align}
\mathcal{J}_{ij}^{0, \alpha} (\mu, T)
    & = - \frac{1}{2 V} \sum_{n, \bk} m_{n \bk, i j}^{\alpha} f_{\mathrm{FD}} (\epsilon_{n \bk})
        + \frac{1}{V} \sum_{n, \bk} \Omega_{n \bk, i j}^{\alpha} g_{n \bk}
\label{eq:clean_limit_J0}
,\end{align}
where $g_{n \bk} = g_{n \bk} (\mu, T) = - \kB T \ln ( 1 + e^{- (\epsilon_{n \bk} - \mu) / \kB T} )$ is the grand-canonical free energy for fermions at energy $\epsilon_{n \bk}$ (see Appendix A for the calculation detail).
From Eqs.~(\ref{eq:clean_limit_L_tilde}) and (\ref{eq:clean_limit_J0}), we then obtain
\begin{align}
L_{Q, i j}^{\alpha}
    & =\tilde{L}_{i j}^{\alpha} + \mathcal{J}_{ij}^{0, \alpha}
    = - \frac{T}{V} \sum_{n, \bk} \Omega_{n \bk, i j}^{\alpha} s_{n \bk}
,\end{align}
where we introduce the entropy $s_{n \bk}$ as
\begin{align}
T s_{n \bk}
    & = (\epsilon_{n \bk} - \mu) f_{\mathrm{FD}} (\epsilon_{n \bk})
\notag \\ & \hspace{1em}
        + \kB T \ln \left[ 1 + \exp \left(- \frac{\epsilon_{n \bk} - \mu}{\kB T}\right) \right]
\label{eq:Snk}
\\
    & = e_{n \bk} - \left( \mu f_{\mathrm{FD}} (\epsilon_{n \bk}) + g_{n \bk} \right)
\\
    & = e_{n \bk} - f_{n \bk}
,\end{align}
where $e_{n \bk} = \epsilon_{n \bk} f_{\mathrm{FD}} (\epsilon_{n \bk})$ is the energy, and $f_{n \bk} = g_{n \bk} + \mu f_{\mathrm{FD}}$ is the Helmholtz free energy.
As already pointed out for the thermal spin torques in Ref.~\cite{kohno2016}, once introducing a thermodynamic quantity,
\begin{align}
\Psi_{i j}^{\alpha} (\mu, T)
    & = - \frac{1}{V} \sum_{n \bk} \Omega_{n \bk, i j}^{\alpha} g_{n \bk}
,\end{align}
and using the relation $f_{\mathrm{FD}} = - \partial g_{n \bk} / \partial \mu$ and $s_{n \bk} = - \partial g_{n \bk} / \partial T$, we have the spin Hall conductivity and spin Nernst coefficient as
\begin{align}
L_{c, i j}^{\alpha}
    & = - \frac{\partial}{\partial \mu} \Psi_{i j}^{\alpha} (\mu, T)
, \\
L_{Q, i j}^{\alpha}
    & = - T \frac{\partial}{\partial T} \Psi_{i j}^{\alpha} (\mu, T)
,\end{align}
which is the spin-current version of the St\v{r}eda formula~\cite{streda1982}. 

To see the convergence at low temperatures, we rewrite the second term in Eq.~(\ref{eq:Snk}) using Eq.~(\ref{eq:NiceTrick}). 
Then, we obtain
\begin{align}
T s_{n \bk}
	& = -\int_{\epsilon_{n \bk} - \mu}^{\infty} \dd{\epsilon'} \epsilon' \frac{\partial f}{\partial \epsilon'} \nonumber \\
    & = \int_{-\infty}^{\epsilon_{n \bk}} \dd{\epsilon} (\epsilon-\mu) \frac{\partial f_{\rm FD}}{\partial \epsilon}
,\end{align}
where we have used the fact that $\epsilon' \frac{\partial f}{\partial \epsilon'}$ is an odd function of $\epsilon'$. 
Thus $T s_{n \bk} \propto T^2 $ as $T \to 0$, indicating that 
the local equilibrium current successfully removes the unphysical divergence of the non-equilibrium spin current in the clean limit.

\subsection{\label{sec:sub:Mott}Genelarization of Mott's formula}
We next establish the spin-current version of Mott's formula, which associates the spin Nernst effect with the spin Hall effect.
To establish this formula, we evaluate the spin Hall conductivity $L_{c, ij}^{\alpha}$ in Eq.~(\ref{eq:response}).
According to the Kubo formula, the spin Hall conductivity is expressed as
\begin{align}
L_{c, i j}^{\alpha}
	& = \lim_{\omega \to 0} \frac{K_{ij}^{\alpha} (\omega) - K_{ij}^{\alpha} (0) }{\zi \omega}
,\end{align}
where $K_{ij}^{\alpha} (\omega)$ is evaluated from the corresponding correlation function in the Matsubara formalism; $K_{ij}^{\alpha} (\omega) = \tilde{K}_{ij}^{\alpha} (\hbar \omega + \zi 0)$ with
\begin{align}
\tilde{K}_{ij}^{\alpha} (\zi \omega_{\lambda})
	& = \int_0^{\beta} \dd{\tau} e^{\zi \omega_{\lambda} \tau} \langle \mathrm{T}_{\tau} j_{\mathrm{s}, i}^{\alpha} (\tau) j_{c, j} \rangle
\label{eq:K_thermal}
,\end{align}
where $j_{c, j} = - e \sum_{\bk} c^{\dagger} v_j (\bk) c^{}_{\bk}$ is the electric current.
With the Matsubara Green function, we write $\tilde{K}_{ij}^{\alpha} (\zi \omega_{\lambda})$ as
\begin{align}
\tilde{K}_{ij}^{\alpha} (\zi \omega_{\lambda})
	& = \frac{e \hbar}{2 \beta V^2} \sum_n \sum_{\bk, \bk'} \tr \left[
		v_{\mathrm{s}, i}^{\alpha} (\bk) \mathcal{G}_{\bk, \bk'} (\zi \epsilon_n^{+})
\notag \right. \\ & \hspace{3em} \left. \times 
		v_j (\bk') \mathcal{G}_{\bk', \bk} (\zi \epsilon_n)
	\right]
.\end{align}
We take the statistical average on the impurity configuration and then obtain
\begin{align}
\langle \tilde{K}_{i j}^{\alpha} (\zi \omega_{\lambda}) \rangle_{\mathrm{av}}
	& = \frac{e \hbar}{2 \beta V} \sum_n \sum_{\bk} \tr \left[
		\Lambda_{\mathrm{s}, i}^{\alpha} (\bk; \zi \epsilon_n, \zi \epsilon_n^{+}) G_{\bk} (\zi \epsilon_n^{+})
\notag \right. \\ & \hspace{3em} \left. \times 
		v_j (\bk) G_{\bk} (\zi \epsilon_n)
	\right]
.\end{align}
Rewriting the Matsubara summation using the contour integral and then taking the analytic continuation $\zi \omega_{\lambda} \to \hbar \omega + \zi 0$, we find
\begin{align}
K_{ij}^{\alpha} (\omega)
	& = - \frac{e \hbar}{2} \int_{-\infty}^{\infty} \frac{\dd{\epsilon}}{2 \pi \zi} \left[
		\bigl( f (\epsilon_{+}) - f (\epsilon_{-}) \bigr) \varphi_{i j}^{\R \A, \alpha} (\epsilon_{+}, \epsilon_{-})
\right. \notag \\ & \hspace{2em} \left.
		+ f (\epsilon_{-}) \varphi_{i j}^{\R\R, \alpha} (\epsilon_{+}, \epsilon_{-})
		- f (\epsilon_{+}) \varphi_{i j}^{\A\A, \alpha} (\epsilon_{+}, \epsilon_{-})
	\right]
,\end{align}
As a result, we obtain the expression of the spin Hall conductivity
\begin{align}
L_{c, ij}^{\alpha}
	& = - \frac{e \hbar^2}{4 \pi} \int_{-\infty}^{\infty} \dd{\epsilon}
		\left[ - \frac{\partial f(\epsilon)}{\partial \epsilon} 
			F_{1, ij}^{\alpha} (\epsilon+\mu)
			+ f(\epsilon) F_{2, ij}^{\alpha} (\epsilon+\mu)
	\right] \nonumber \\
	& = - \frac{e \hbar^2}{4 \pi} \int_{-\infty}^{\infty} \dd{\epsilon}
		\left( - \frac{\partial f_{\mathrm{FD}}}{\partial \epsilon} \right)
		\left[
			F_{1, ij}^{\alpha} (\epsilon)
			+ \mathcal{F}_{2, ij}^{\alpha} (\epsilon)
	\right]
\label{eq:Lc}
,\end{align}
where we have carried out an integration by parts in the second term.
From Eq.~(\ref{eq:precise_LQ}) and Eq.~(\ref{eq:Lc}), we obtain the following spin-current version of Mott's formula at low temperatures
\begin{align}
L_{Q, y x}^z
	& = \frac{\pi^2 (\kB T)^2}{- 3 e} \frac{\partial \sigma_{\mathrm{SH}} (\eF)}{\partial \eF}
\label{eq:Mott}
,\end{align}
where $\sigma_{\mathrm{SH}} (\eF)$ is the spin Hall conductivity at the zero temperature given as
\begin{align}
\sigma_{\mathrm{SH}} (\eF)
	& = L_{c, y x}^z \big|_{T = 0}
	= - \frac{e \hbar^2}{4 \pi} 
		\left[
			F_{1, y x}^{z} (\eF)
			+ \mathcal{F}_{2, y x}^{z} (\eF)
	\right]
.\end{align}

\section{\label{sec:3DDirac}Dirac electron system}
\subsection{Model}
As a demonstration of our theory, we compute the spin Nernst coefficient of a three-dimensional Dirac electron system.
Following Ref.~\cite{fuseya2012}, we consider the effective (isotropic) Dirac Hamiltonian,
\begin{align}
\mathcal{H}_{\mathrm{D}}
	& = \begin{pmatrix}
		\Delta
	&	\zi \hbar v \bk \cdot \bm{\sigma}
	\\ -\zi \hbar v \bk \cdot \bm{\sigma}
	&	- \Delta 
	\end{pmatrix}
	= - \hbar v \rho_2 \bk \cdot \bm{\sigma}
		+ \Delta \rho_3
,\end{align}
where $v$ is the velocity, $\bm{\sigma} = (\sigma^x, \sigma^y, \sigma^z)$ are the Pauli matrices in spin space, and $\rho_i$ ($i = 1, 2, 3$) are the Pauli matrices in particle-hole space.
The eigenenergies of the Dirac Hamiltonian are given as $\pm \epsilon_k  = \pm \sqrt{\hbar^2 v^2 k^2 + \Delta^2}$.
We use $\rho_0$ and $\sigma^0$ as the unit matrices when emphasizing them.
We consider the short-range impurity potential $V (\br) = u \sum_{i} \rho_0 \sigma^0 \delta (\br - \bm{R}_i)$, where $u$ is the impurity potential strength and $\bm{R}_i$ is the positions of the impurities.
The retarded/advanced Green function within the Born approximation is given by Eq.~(2.10) with Eqs.~(2.11) in Ref.~\cite{fukazawa2017}.
Note that the treatment of impurities in this model is described in Ref.~\cite{fujimoto2022a} and \cite{fukazawa2017}.

The velocity operator in the Dirac electron system is given by
\begin{align}
v_i
	& = \frac{1}{\hbar} \frac{\partial \mathcal{H}_{\mathrm{D}}}{\partial k_i}
	= - v \rho_2 \sigma^i
\qquad
	(i = x, y, z)
,\end{align}
and the velocity operator of the spin current with spin component $\alpha$ is given by
\begin{align}
v_{\mathrm{s}, i}^{\alpha}
	& = \frac{1}{2 \mu_{\mathrm{B}}} [ v_i, \mu_{\mathrm{s}, \alpha} ]_{+}
\qquad (i, \alpha = x, y, z)
\\
	& = - \frac{g^{*} v}{2} \epsilon_{i \alpha j} \rho_1 \sigma^j
,\end{align}
where $[ A, B ]_{+} = A B + B A$ is the anticommutator and $\bm{\mu}_{\mathrm{s}} = - (g^{*} \mu_{\mathrm{B}} / 2) \rho_3 \bm{\sigma}$ is the spin magnetic moment with the effective $g$-factor $g^{*} = 2 m v^2 / \Delta$ and the Bohr magneton $\mu_{\mathrm{B}}$.
Using these, the electric, spin, and heat current operators in the second quantization are given by
\begin{align}
J_i
	& = e v \sum_{\bk} c_{\bk}^{\dagger} \rho_2 \sigma^i c_{\bk},\\
j_{\mathrm{s}, i}^{\alpha}
	& = - \frac{g^{*} v}{2 V} \epsilon_{i \alpha j} \sum_{\bk} c_{\bk}^{\dagger} \rho_1 \sigma^j c_{\bk},\\
J_{\mathrm{Q}, i}
	& = - \frac{v}{2} \sum_{\bk} \left(
		\dot{c}_{\bk}^{\dagger} \rho_2 \sigma^i c_{\bk}
		- c_{\bk}^{\dagger} \rho_2 \sigma^i \dot{c}_{\bk}
	\right)
,\end{align}
where $V$ is the system volume.

The retarded Green function with the Born approximation is given by~\cite{fukazawa2017,fujimoto2022a}
\begin{align}
G^{\mathrm{R}}_{\bm{k}} (\epsilon)
	& = \frac{1}{ D^{\mathrm{R}}_{\bm{k}} (\epsilon+\mu) }
		\Bigl( g^{\mathrm{R}}_0 (\epsilon+\mu) + \rho_2 \bm{g}^{\mathrm{R}}_2 (\bm{k}) \cdot \bm{\sigma} + \rho_3 g^{\mathrm{R}}_3 (\epsilon+\mu) \Bigr)
\label{eq:retarded_G_def}
,\end{align}
where
\begin{subequations}
\begin{align}
D^{\mathrm{R}}_{\bm{k}} (\epsilon)
	& = ( \epsilon + i \gamma_0 (\epsilon) )^2
	- \hbar^2 v^2 k^2
	- (\Delta - i \gamma_3 (\epsilon))^2
\label{eq:DR_k}
, \\
g^{\mathrm{R}}_0 (\epsilon)
	& = \epsilon + i \gamma_0 (\epsilon)
, \\
\bm{g}^{\mathrm{R}}_2 (\bm{k})
	& = - \hbar v \bm{k}
, \\
g^{\mathrm{R}}_3 (\epsilon)
	& = \Delta - i \gamma_3 (\epsilon)
.\end{align}
\end{subequations}
Here,
\begin{subequations}
\begin{align}
\gamma_0 (\epsilon)
	& = \frac{\pi}{2} n_{\mathrm{i}} u^2 \nu (\epsilon)
, \\
\gamma_3 (\epsilon)
	& = \frac{\pi}{2} n_{\mathrm{i}} u^2 \frac{\Delta}{\epsilon} \nu (\epsilon)
\end{align}
\label{eq:damping-constants}%
\end{subequations}
are the damping rates of electrons with the impurity concentration $n_{\mathrm{i}}$.
$\nu (\epsilon)$ is the density of states given by
\begin{align}
\nu (\epsilon)
	& = \frac{1}{V} \sum_{\bm{k}, \eta = \pm} \delta (\epsilon - \eta \varepsilon_k)
	= \frac{|\epsilon|}{2 \pi^2 \hbar^3 v^3} \sqrt{\epsilon^2 - \Delta^2} \sum_{\eta} \Theta ( \eta \epsilon - \Delta )
\label{eq:DOS}
.\end{align}
The advanced Green function $G^{\A}_{\bk} (\epsilon)$ is obtained by replacing $\gamma_0$ and $\gamma_3$ with $- \gamma_0$ and $- \gamma_3$ in $G^{\R}_{\bk} (\epsilon)$.
In this model, the Fourier coefficient of the impurity potential is $u (\bq) = u$ because the short-range impurity potential is assumed.

\subsection{\label{sec:sub:response}Response coefficients}
We consider that temperature gradients are applied along $x$ direction and the response of the spin current flowing in the $y$ direction with the $z$ spin polarization; $i = y, j = x, \alpha = z$.
We evaluate the leading order with respect to the $\mu \tau / \hbar$ and hence we approximate Eq.~(\ref{eq:F1}) as
\begin{align}
F_{1, y x}^z (\epsilon + \mu)
	& \simeq \varphi_{y x}^{\R \A, z} (\epsilon, \epsilon)
	\equiv \varphi_{\mathrm{ladder}}
		+ \varphi_{\mathrm{skew}}
,\end{align}
where $\varphi_{\mathrm{ladder}}$ is the contribution of the side jump, and $\varphi_{\mathrm{skew}}$ is the contribution of the skew scattering, which are given by
\begin{align}
\varphi_{\mathrm{ladder}}
	& = \frac{1}{4 V} \sum_{\bk} \tr \left[
		\rho_1 \sigma^x G^{\R}_{\bk} \Lambda_{2, x} G^{\A}_{\bk}
	\right]
, \\
\varphi_{\mathrm{skew}}
	& = \frac{n_{\mathrm{i}} u^3}{4 V^3} \sum_{\bk, \bk', \bk''} \tr \left[
		\Lambda_{1, x}^{*} G^{\R}_{\bk}
		G^{\R}_{\bk'} \Lambda_{2, x} G^{\A}_{\bk'}
		G^{\A}_{\bk''}
		G^{\A}_{\bk}
\right. \notag \\ & \hspace{6em} \left.
        + \Lambda_{1, x}^{*} G^{\R}_{\bk}
		G^{\R}_{\bk''}
		G^{\R}_{\bk'} \Lambda_{2, x} G^{\A}_{\bk'}
		G^{\A}_{\bk}
	\right]
,\end{align}
with $G^{\R/\A}_{\bk} = G^{\R/\A}_{\bk} (\epsilon)$.
$\Lambda_{1, x}^{*}$ and $\Lambda_{2,x}$ are velocity vertexes of the spin and charge currents with the ladder-type vertex corrections satisfying
\begin{align}
\Lambda_{1, x}^{*}
	& = \rho_1 \sigma^x
		+ \frac{n_{\mathrm{i}} u^2}{V} \sum_{\bk} G^{\A}_{\bk} \Lambda_{1, x}^{*} G^{\R}_{\bk}
, \\
\Lambda_{2, x}
	& = \rho_2 \sigma^x
		+ \frac{n_{\mathrm{i}} u^2}{V} \sum_{\bk} G^{\R}_{\bk} \Lambda_{2, x} G^{\A}_{\bk}
.\end{align}
From the calculation of Ref.~\cite{fukazawa2017}, we have
\begin{align}
\Lambda_{1, x}^{*}
	& = \frac{1}{1 - 2 U} \rho_1 \sigma^x
	- \frac{V}{(1 - U) (1 - 2 U)} \rho_2 \sigma^x
, \\
\Lambda_{2, x}
	& = - \frac{V}{(1 - U) (1 - 2 U)} \rho_1 \sigma^x
	+ \frac{1}{1 - U} \rho_2 \sigma^x
\end{align}
with
\begin{align}
U (\epsilon)
	= \frac{1}{3} \frac{\epsilon^2 - \Delta^2}{\epsilon^2 + \Delta^2}
, \qquad
V (\epsilon)
	= n_{\mathrm{i}} u^2 \frac{\pi \Delta}{\epsilon^2 + \Delta^2}
.\end{align}
Immediately, we find
\begin{align}
\varphi_{\mathrm{ladder}}
	& = - \frac{1}{n_{\mathrm{i}} u^2} \frac{V}{(1 - U) (1 - 2 U)},\\
\varphi_{\mathrm{skew}}
	& = \frac{1}{n_{\mathrm{i}} u} \frac{2 U}{1 - 2 U} \frac{U}{1 - U}
		\frac{- 2 \gamma_3}{n_{\mathrm{i}} u^2}
.\end{align}

Eq.~(\ref{eq:F2}) is calculated as
\begin{align}
F_{2, ij}^{\alpha} (\epsilon + \mu)
	& = \varphi_{\mathrm{sea}} (\epsilon + \mu)
\end{align}
with
\begin{align}
\varphi_{\mathrm{sea}} (\epsilon + \mu)
	& = \frac{1}{8 \Omega} \sum_{\bk} \tr \left[
		\rho_1 \sigma^x G^{\R}_{\bk} \rho_2 \sigma^x \left( \partial_{\epsilon} G^{\R}_{\bk} \right)
\right. \notag \\ & \hspace{-1em} \left.
		- \rho_1 \sigma^x \left( \partial_{\epsilon} G^{\R}_{\bk} \right) \rho_2 \sigma^x G^{\R}_{\bk}
		- (\R \to \A)
	\right]
.\end{align}
The leading order in $\varphi_{\mathrm{sea}}(\epsilon)$ for $\mu \tau/\hbar$ is zero so that the limit $n_{\mathrm{i}} \to 0$ can be taken.
Using
\begin{align}
\partial_{\epsilon} G^{\R/\A}_{\bk}
	& = \frac{1}{D^{\R}_{\bk} (\epsilon)}
		- \frac{\partial_{\epsilon} D^{\R}_{\bk} (\epsilon)}{D^{\R}_{\bk} (\epsilon)} G^{\R}_{\bk} (\epsilon)
,\end{align}
we obtain
\begin{align}
\frac{1}{8}
	& \tr \left[
		\rho_1 \sigma^x G^{\R}_{\bk} \rho_2 \sigma^x \left( \partial_{\epsilon} G^{\R}_{\bk} \right)
\right. \notag \\ & \hspace{3em} \left.
        - \rho_1 \sigma^x \left( \partial_{\epsilon} G^{\R}_{\bk} \right) \rho_2 \sigma^x G^{\R}_{\bk}
		- (\R \to \A)
\right]
\notag \\
	& = \frac{1}{8 D^{\R}_{\bk} (\epsilon)} \tr \left[
		\rho_1 \sigma^x G^{\R}_{\bk} \rho_2 \sigma^x
		- \rho_1 \sigma^x \rho_2 \sigma^x G^{\R}_{\bk}
		- (\R \to \A)
\right]
\notag \\
	& = - \frac{\zi}{4 D^{\R}_{\bk} (\epsilon)} \tr \left[
		\rho_3 G^{\R}_{\bk}
		- \rho_3 G^{\A}_{\bk}
\right]
\notag \\
	& = - \zi \Delta \left( \frac{1}{\{ D^{\R}_{\bk} (\epsilon) \}^2} - \frac{1}{\{ D^{\A}_{\bk} (\epsilon) \}^2} \right)
\label{eq:phi_sea}
.\end{align}
From Ref.~\cite{fukazawa2017}, Eq.~(\ref{eq:phi_sea}) is evaluated as
\begin{align}
\frac{1}{8} \tr \left[ \cdots \right]
	& = \frac{\pi \Delta}{2} \sum_{\eta = \pm} \frac{1}{\epsilon + \mu}
		\frac{\partial}{\partial \epsilon} \left( \frac{ \delta (\epsilon + \mu - \eta \epsilon_k) }{\epsilon + \mu} \right).
\end{align}
We thus have
\begin{align}
\varphi_{\mathrm{sea}} (\epsilon)
	& = \frac{\pi \Delta}{2 V} \sum_{\bk} \sum_{\eta = \pm} \frac{1}{\epsilon}
		\frac{\partial}{\partial \epsilon} \left( \frac{ \delta (\epsilon - \eta \epsilon_k) }{\epsilon} \right)
\notag \\
	& = \frac{\pi \Delta}{2 \epsilon}
		\frac{\partial}{\partial \epsilon} \left( \frac{ \nu (\epsilon) }{\epsilon} \right)
\notag \\
	& = \frac{\partial}{\partial \epsilon} \left(
		\frac{\pi \Delta}{2 \epsilon^2} \nu (\epsilon)
	\right)
	+ \frac{\pi \Delta}{2 \epsilon^3} \nu (\epsilon)
\label{eq:varphi_sea}
.\end{align}
We finally obtain
\begin{align}
& \int_{-\infty}^{\infty} \mathrm{d}\epsilon f (\epsilon) \epsilon \varphi_{\mathrm{sea}} (\epsilon + \mu)
\notag \\
	& = \int_{-\infty}^{\infty} \mathrm{d}\epsilon \epsilon \left\{
		\left( - \frac{\partial f}{\partial \epsilon} \right)
			\frac{\pi \Delta}{2 (\epsilon  + \mu)^2} \nu (\epsilon + \mu)
\right. \notag \\ & \hspace{10em} \left.
            + f (\epsilon) \frac{\pi \Delta}{2 (\epsilon + \mu)^3} \nu (\epsilon + \mu)
	\right\}
\\
	& = \int_{-\infty}^{\infty} \mathrm{d}\epsilon (\epsilon - \mu) \left\{
		\left( - \frac{\partial f_{\mathrm{FD}}}{\partial \epsilon} \right)
			\frac{\pi \Delta}{2 \epsilon^2} \nu (\epsilon)
		+ f_{\mathrm{FD}} (\epsilon) \frac{\pi \Delta}{2 \epsilon^3} \nu (\epsilon)
	\right\}
 \label{eq:phi_sea1}
.\end{align}
Note that the last term in Eq.~(\ref{eq:phi_sea1}) causes the divergence of the non-equilibrium spin current at $T =0$.

\subsection{\label{sec:sub:localequilibrium}Local equilibrium current}
To obtain the proper spin Nernst coeffient, we compute the local equilibrium current.
In this model, Eq.~(\ref{eq:Phi_ij}) is computed as
\begin{align}
& \Phi_{yx}^z (\epsilon + \mu)
\notag \\
	& = - \hbar \, \Im \sum_{\bk} \tr \left[
		\left( - \frac{g^{*} v}{2 V} \epsilon^{y z j} \rho_1 \sigma^j \right) G^{\R}_{\bk}
			\left( - v \rho_2 \sigma^x \right) \left( \partial_{\epsilon} G^{\R}_{\bk} \right)
\right. \notag \\ & \hspace{1em} \left.
            - \left( - \frac{g^{*} v}{2 V} \epsilon^{y z j} \rho_1 \sigma^j \right) \left( \partial_{\epsilon} G^{\R}_{\bk} \right)
			\left( - v \rho_2 \sigma^x \right) G^{\R}_{\bk}
	\right]
\notag \\
	& = - \frac{\hbar g^{*} v^2}{2 V} \Im \sum_{\bk} \tr \left[
		\rho_1 \sigma^x G^{\R}_{\bk}
		\rho_2 \sigma^x \left( \partial_{\epsilon} G^{\R}_{\bk} \right)
\right. \notag \\ & \hspace{1em} \left.
		- \rho_1 \sigma^x \left( \partial_{\epsilon} G^{\R}_{\bk} \right)
		\rho_2 \sigma^x G^{\R}_{\bk}
	\right]
\notag \\
	& = 2 \zi \hbar g^{*} v^2 \varphi_{\mathrm{sea}}
.\end{align}
We then obtain the local equilibrium current 
\begin{align}
\mathcal{J}_{y x}^{0, z} (\mu, T)
	& = \frac{\hbar g^{*} v^2}{\pi} \int_{-\infty}^{\infty} \mathrm{d}\epsilon f (\epsilon)
		\int_{-\infty}^{\mu} \mathrm{d}\mu' \varphi_{\mathrm{sea}} (\epsilon + \mu)
.\end{align}

In total, we obtain the spin Nernst coefficient
\begin{align}
L_{Q, y x}^{z}&=\tilde{L}_{Q, y x}^{z} + \mathcal{J}_{y x}^{0, z}\nonumber\\
	& = \frac{\hbar g^{*} v^2}{\pi} \int_{-\infty}^{\infty} \mathrm{d}\epsilon
		\left[
			\left( - \frac{\partial f}{\partial \epsilon} \right) \epsilon \varphi_{\mathrm{surf}} (\epsilon + \mu)
\right. \notag \\ & \hspace{-1em} \left.
			+ f (\epsilon) \epsilon \varphi_{\mathrm{sea}} (\epsilon + \mu)
			+ f (\epsilon) \int_{-\infty}^{\mu} \mathrm{d}\mu' \varphi_{\mathrm{sea}} (\epsilon + \mu')
		\right]
.\end{align}
The obtained spin Nernst coefficient does not involve the unphysical divergence at $T=0$.
Below, we clarify that the absence of the unphysical divergence of the computed spin Nernst effect.
Because $\varphi_{\mathrm{sea}}$ is the function of $\epsilon + \mu'$, we have,
\begin{align}
\Phi_{\mathrm{sea}} (\epsilon + \mu)\equiv \int_{-\infty}^{\mu} \mathrm{d}\mu' \varphi_{\mathrm{sea}}
	= \int_{-\infty}^{\epsilon} \mathrm{d}\epsilon' \varphi_{\mathrm{sea}} (\epsilon' + \mu)
.\end{align}
With $\frac{\partial}{\partial \epsilon} \Phi_{\mathrm{sea}} = \varphi_{\mathrm{sea}}$, we obtain
\begin{align}
\int_{-\infty}^{\infty} \mathrm{d}\epsilon
& \left(
	f (\epsilon) \epsilon \varphi_{\mathrm{sea}} (\epsilon + \mu)
	+ f (\epsilon) \int_{-\infty}^{\mu} \mathrm{d}\mu' \varphi_{\mathrm{sea}} (\epsilon + \mu)
\right)
\notag \\
	& = \int_{-\infty}^{\infty} \mathrm{d}\epsilon \left(
		f (\epsilon) \epsilon \frac{\partial}{\partial \epsilon} \Phi_{\mathrm{sea}} (\epsilon + \mu)
		+ f (\epsilon) \Phi_{\mathrm{sea}} (\epsilon + \mu)
	\right)
\notag \\
	& = \int_{-\infty}^{\infty} \mathrm{d}\epsilon
		f (\epsilon) \frac{\partial}{\partial \epsilon} \left( \epsilon \Phi_{\mathrm{sea}} (\epsilon + \mu) \right)
\notag \\
	& = \int_{-\infty}^{\infty} \mathrm{d}\epsilon
		\left( - \frac{\partial f_{\mathrm{FD}}}{\partial \epsilon} \right) (\epsilon - \mu) \Phi_{\mathrm{sea}} (\epsilon)
.\end{align}
We finally obtain
\begin{align}
L_{Q, y x}^{z}
	& = \tilde{L}_{Q, y x}^{z} + \mathcal{J}_{y x}^{0, z}
\notag \\
	& = \frac{\hbar g^{*} v^2}{\pi} \int_{-\infty}^{\infty} \mathrm{d}\epsilon
			\left( - \frac{\partial f_{\mathrm{FD}}}{\partial \epsilon} \right) (\epsilon - \mu)
				\left\{ \varphi_{\mathrm{surf}} (\epsilon) + \Phi_{\mathrm{sea}} (\epsilon) \right\}
,
\label{eq:SNC}
\end{align}
Eq.~(\ref{eq:SNC}) clarifies that the calculated spin Nernst conductivity goes to zero as $T\to 0$ and does not involve the unphysical divergence.

From Eq.~(\ref{eq:varphi_sea}), we express the local equilibrium current as
\begin{align}
\mathcal{J}_{y x}^{0, z} (\mu, T)
	& = \frac{\hbar g^{*} v^2}{\pi} \int_{-\infty}^{\infty} \mathrm{d}\epsilon f (\epsilon)
		\Phi_{\mathrm{sea}} (\epsilon + \mu)
,\end{align}
where
\begin{align}
\Phi_{\mathrm{sea}} (\epsilon)
	& = \int_{-\infty}^{\epsilon} \mathrm{d}\epsilon' \varphi_{\mathrm{sea}} (\epsilon')
	= \frac{\pi \Delta}{2 \epsilon^2} \nu (\epsilon)
	+ \pi \Delta \int_{-\infty}^{\epsilon} \mathrm{d}\epsilon' \frac{\nu (\epsilon')}{2 \epsilon'^3}
.\end{align}

\subsection{Results and Discussion}
We finally summarize the results and discuss the spin Nernst effect in the three-dimensional Dirac electron system.
When the electric field and temperature gradient are applied along the $x$ direction, the response of the spin current flowing in the $y$ direction with the $z$ spin polarization is described by
\begin{align}
j_{\mathrm{s}, y}^{z}
	& = L_{c, y x}^z E_x + L_{Q, y x}^z \left( - \frac{\nabla_x T}{T} \right)
.\end{align}
Note that the other components of the spin current response are zero.
From Ref.~\cite{fukazawa2017} and Eq.~(\ref{eq:SNC}), the spin Hall conductivity and the spin Nernst coefficient are expressed as
\begin{align}
L_{c, y x}^{z}
	& = \frac{- e \hbar g^{*} v^2}{\pi} \int_{-\infty}^{\infty} \mathrm{d}\epsilon
			\left( - \frac{\partial f_{\mathrm{FD}}}{\partial \epsilon} \right)
				\left\{ \varphi_{\mathrm{surf}} (\epsilon) + \Phi_{\mathrm{sea}} (\epsilon) \right\}
, \\
L_{Q, y x}^{z}
	& = \frac{\hbar g^{*} v^2}{\pi} \int_{-\infty}^{\infty} \mathrm{d}\epsilon
			\left( - \frac{\partial f_{\mathrm{FD}}}{\partial \epsilon} \right) (\epsilon - \mu)
				\left\{ \varphi_{\mathrm{surf}} (\epsilon) + \Phi_{\mathrm{sea}} (\epsilon) \right\}
.\end{align}
$\varphi_{\mathrm{surf}} (\epsilon)$ describes the two distinct extrinsic (impurity) contributions to these transport phenomena and expressed as
\begin{align}
\varphi_{\mathrm{surf}} (\epsilon)
	& = \varphi_{\mathrm{ladder}} (\epsilon)
		+ \varphi_{\mathrm{skew}} (\epsilon)
.\end{align}
$\varphi_{\mathrm{ladder}} (\epsilon)$ arises from the ladder-type diagrams and describes the so-called side-jump contribution
\begin{align}
\varphi_{\mathrm{ladder}} (\epsilon)
	& = - \frac{\pi \Delta}{\epsilon^2 + \Delta^2} \beta (\epsilon) \nu (\epsilon)
.\end{align}
and $\varphi_{\mathrm{skew}} (\epsilon)$ describes the skew-scattering contribution
\begin{align}
\varphi_{\mathrm{skew}} (\epsilon)
	& = - \frac{2}{n_{\mathrm{i}} u} \frac{\pi \Delta}{\epsilon} \{ U (\epsilon) \}^2 \beta (\epsilon)
		\nu (\epsilon)
.\end{align}
Here, $\nu (\epsilon)$ is the density of states given by Eq.~(\ref{eq:DOS}) and
\begin{align}
\beta (\epsilon)
	& = \frac{9 (\epsilon^2 + \Delta^2)^2}{ 2 (\epsilon^2 + 2 \Delta^2) (\epsilon^2 + 5 \Delta^2) }
		+ \mathcal{O} (n_{\mathrm{i}}^2)
, \\
U (\epsilon)
	& = \frac{1}{3} \frac{ \epsilon^2 - \Delta^2 }{ \epsilon^2 + \Delta^2 }
		+ \mathcal{O} (n_{\mathrm{i}}^2)
\label{eq:longitudinal_conductivity}
.\end{align}

The intrinsic contribution $\Phi_{\mathrm{sea}} (\epsilon)$ is given by
\begin{align}
\Phi_{\mathrm{sea}} (\epsilon)
	& = \frac{\pi \Delta}{2 \epsilon^2} \nu (\epsilon)
	+ \pi \Delta \int_{-\infty}^{\epsilon} \mathrm{d}\epsilon' \frac{\nu (\epsilon')}{2 \epsilon'^3}
.\end{align}

As shown before, at low temperatures, the spin-current version of Mott's formula is satisfied as
\begin{align}
L_{Q, y x}^z
	& = \frac{\pi^2 (\kB T)^2}{- 3 e} \frac{\partial \sigma_{\mathrm{SH}} (\eF)}{\partial \eF}
,\end{align}
where $\sigma_{\mathrm{SH}} (\eF)$ is the spin Hall conductivity at the zero temperature given as
\begin{align}
\sigma_{\mathrm{SH}} (\eF)
	& = L_{c, y x}^z \big|_{T = 0}
	= \frac{- e \hbar g^{\ast} v^2}{\pi} 
		\left\{ \varphi_{\mathrm{surf}} (\eF) + \Phi_{\mathrm{sea}} (\eF) \right\}
.\end{align}

\begin{figure*}[tbhp]
\centering
\includegraphics[width=\linewidth]{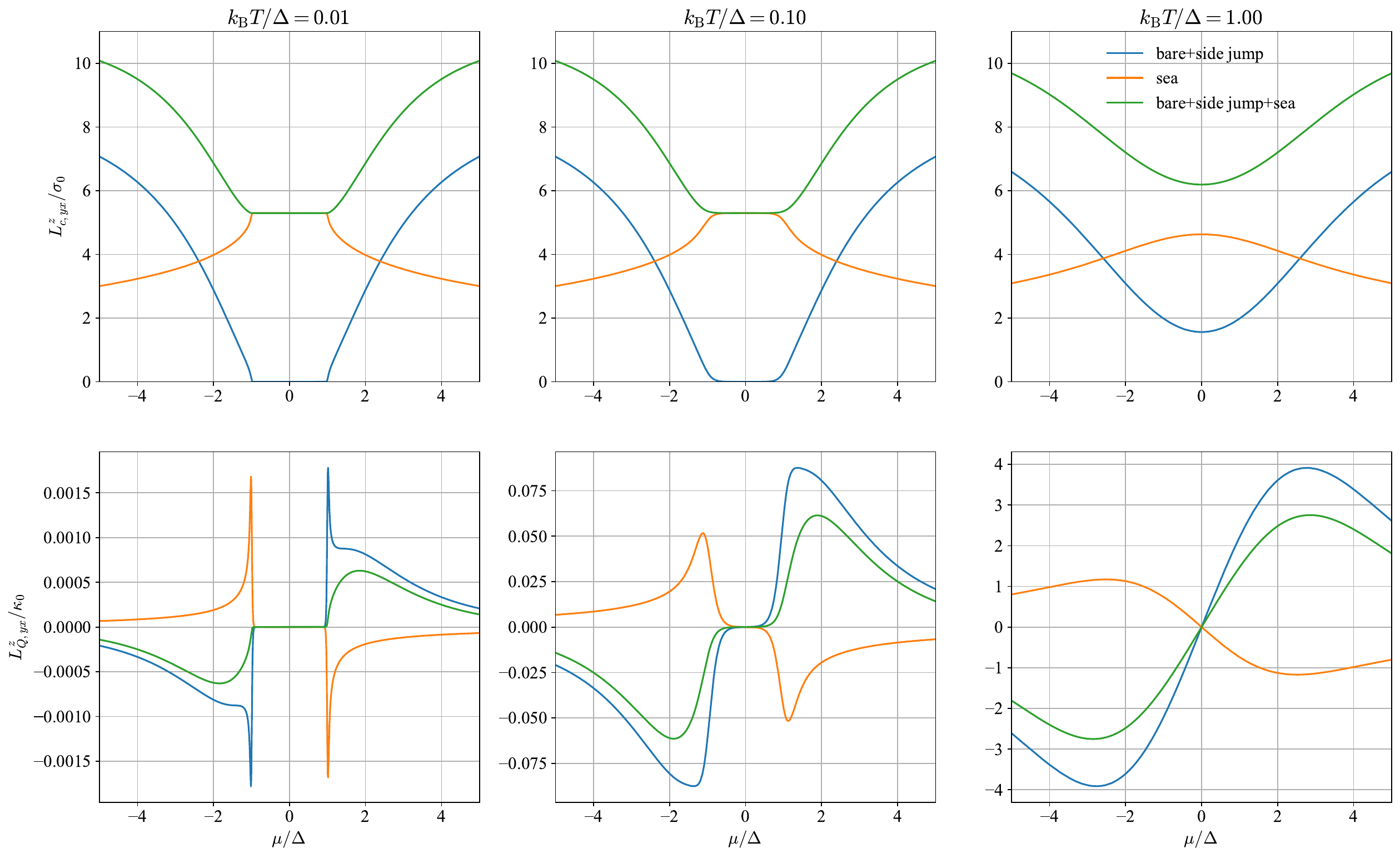}
\caption{\label{fig:1}Chemical potential dependences of the spin Hall and spin Nernst coefficients without the skew-scattering contributions.}
\includegraphics[width=\linewidth]{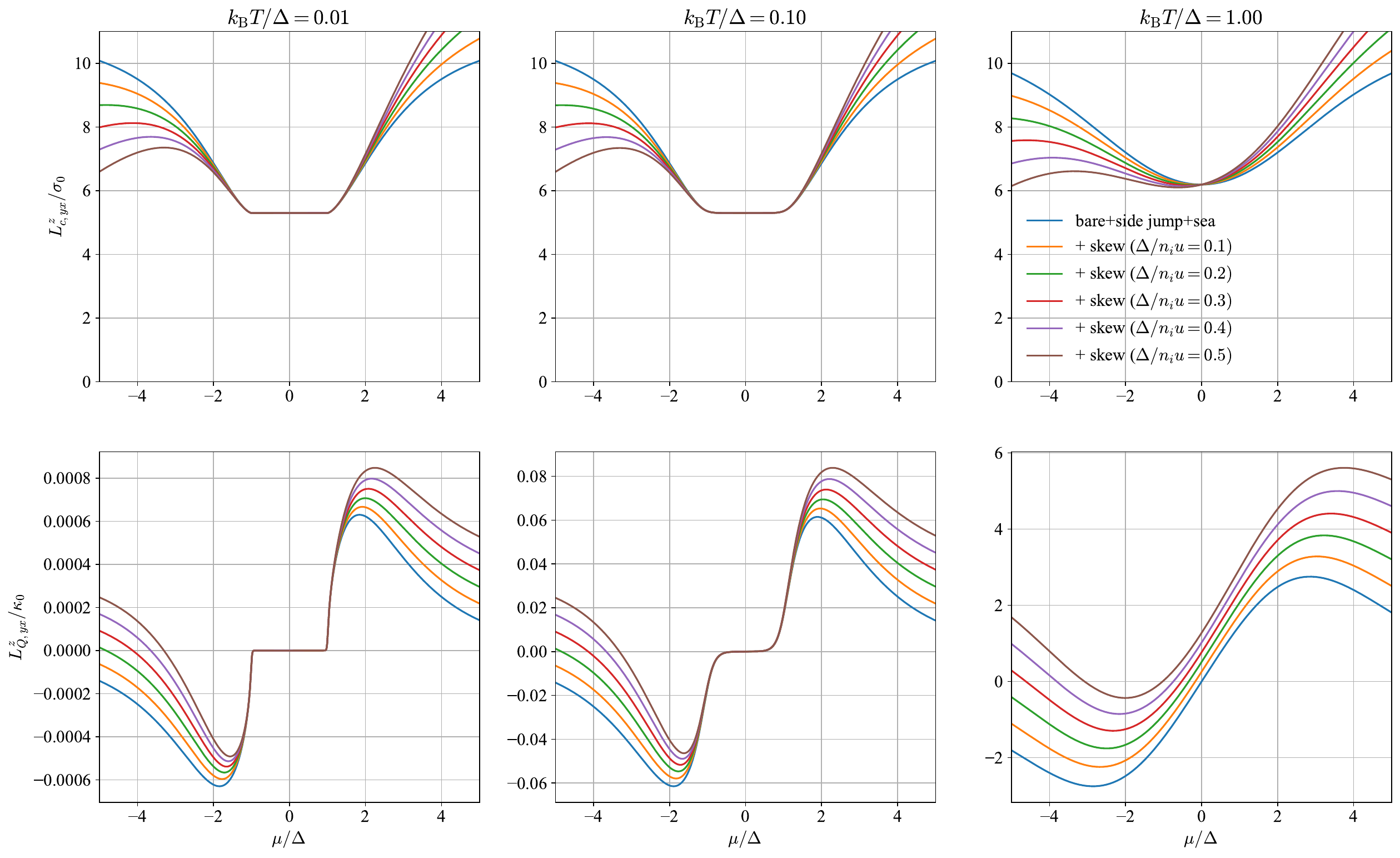}
\caption{\label{fig:2}Chemical potential dependences of the spin Hall and spin Nernst coefficients for various $n_{\mathrm{i}} u / \Delta$.}
\end{figure*}

Now, we discuss the dependence of the spin Hall and spin Nernst coefficient on the chemical potential depicted in Fig.~\ref{fig:1}.
The upper panels of Fig.~\ref{fig:1} show the chemical potential dependence of the spin Hall coefficient for $\kB T / \Delta = 0.01, 0.1$, and $1$, where the skew-scattering contribution is not included.
When the chemical potential lies in the band gap ($|\mu / \Delta| < 1$), the spin Hall coefficient takes the nonzero (and not-quantized) value due to the Fermi-sea term, which arises from the intrinsic contribution.
This feature has been discovered in Ref.~\cite{fuseya2012,fukazawa2017}.
At higher temperatures, the sharp chemical potential dependence becomes masked, although the magnitude of the spin Hall coefficient does not change significantly.

The lower panels of Fig.~\ref{fig:1} indicate the chemical potential dependence of the spin Nernst coefficient for $\kB T / \Delta = 0.01, 0.1$, and $1$ without the skew-scattering contribution.
The spin Nernst coefficient is odd for the chemical potential.
The blue and yellow lines, corresponding to the Fermi-surface and Fermi-sea terms, respectively, are sensitive to the chemical potential, but the sum of them is not.
This is because the contribution from the spin-dependent magnetic moment in the Fermi surface term is canceled by that in the Fermi sea term.
Note that, in contrast with the spin Hall effect, the magnitude of the spin Nernst effect is sensitive to the temperature.

Next, we show the contribution from the skew scattering in Fig.~\ref{fig:2}.
We plot the spin Hall and spin Nernst coefficients in the upper and lower panels, respectively.
The spin Hall coefficient without the skew-scattering contribution is even for the chemical potential, whereas the skew-scattering contribution is odd for it.
Hence, the total spin Hall coefficient is neither even nor odd for the chemical potential.
Similarly, for the spin Nernst coefficient, the skew-scattering contribution is an even function for the chemical potential, and thus the total spin Nernst coefficient has no such symmetry.
We again note that the increase of the temperature leads to the significant enhancement of the spin Nernst coefficient, whereas the spin Hall coefficient does not change drastically.

\begin{figure}[tbhp]
\centering
\includegraphics[width=\linewidth]{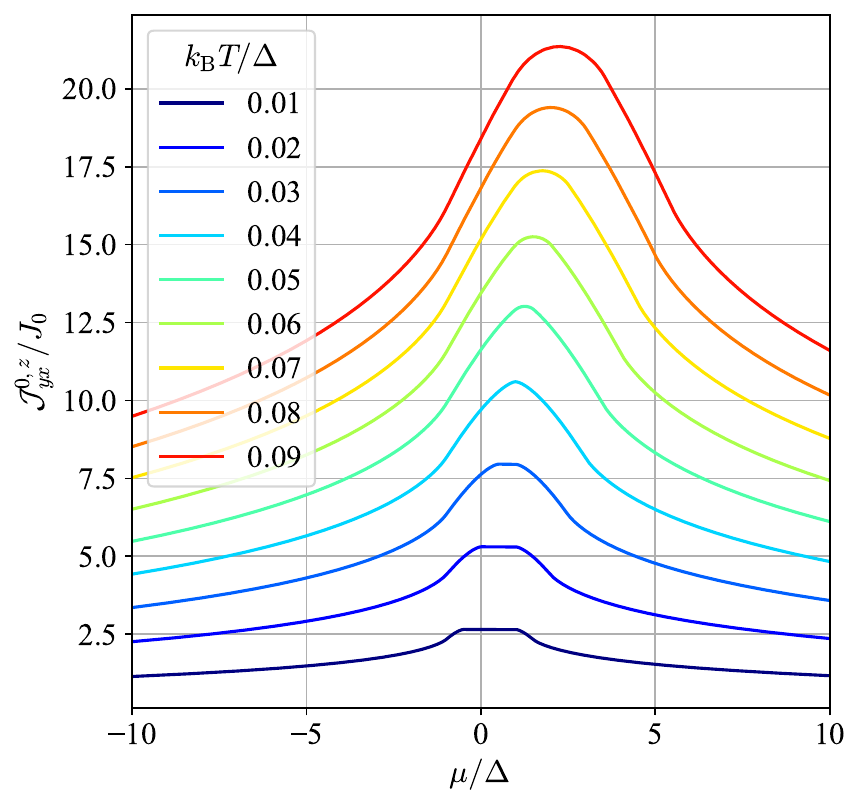}
\caption{\label{fig:3}Chemical potential dependence of the local equilibrium current for various temperatures.}
\end{figure}

Then, we plot the chemical potential dependence of the local equilibrium current for low temperatures in Fig.~\ref{fig:3}.
We find that the dependence for $\Delta / \kB T \sim 0.1$ is different from that of the Fermi-sea term, whereas the dependence for $\Delta / \kB T \sim 0.01$ is similar to that of the Fermi-sea term (see the orange line in the upper panel of Fig.~\ref{fig:1}).

\begin{figure*}[tbhp]
\centering
\includegraphics[width=\linewidth]{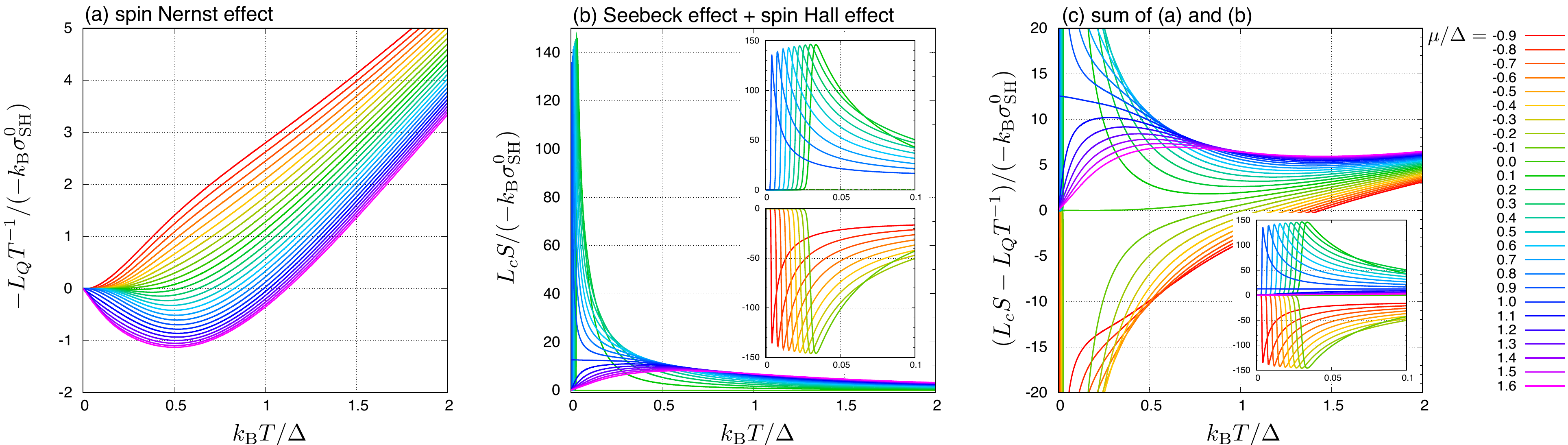}
\caption{\label{fig:4}Temperature dependences of two paths of spin current generation for various chemical potentials.}
\end{figure*}
Finally, we note that temperature gradients induce the transverse spin current in two ways: the spin Nernst effect and the combined effect of the Seebeck effect and spin Hall effect.
Both mechanisms should be considered because the sum of these are observed in experiments.
The Seebeck effect of the three-dimensional Dirac electron system is calculated in Ref.~\cite{fujimoto2022a}.
As shown in Fig.~\ref{fig:4}, when the chemical potential lies near the band bottom, the combined effect is dominant, because the Seebeck effect is enhanced as $1/T$.
This enhancement would be useful for thermoelectric power generation via the thermospin conversion~\cite{fujimoto2022b}.

Here, we discuss previous studies on the spin Nernst effect in the two-dimensional Rashba electron system~\cite{dyrdal2016,dyrdal2016a}.
The results in Ref.~\cite{dyrdal2016,dyrdal2016a} have similarities to ours, which are summarized as follows:
First, the spin Nernst conductivity evaluated using the Kubo formula exhibits physically unrealistic behavior at low temperatures, but by considering the local equilibrium contribution (spin-resolved orbital magnetization), this unphysical behavior is corrected.
Second, the precise spin Nernst conductivity shows a $T^2$ dependence at low temperatures, as demonstrated in Eq.~(\ref{eq:Mott}) and Fig.~\ref{fig:4}~(a) in this paper as well as Fig.~4~of Ref.~\cite{dyrdal2016a}.
However, there are some differences as follows:
The previous work~\cite{dyrdal2016a} is based on the eigenstate representation, i.e., the spin-resolved expression of the Rashba Hamiltonian, while our calculations are performed without using the eigenstate representation and instead rely on the Green function for general Hamiltonian.
The advantage of our approach is that we do not need to analytically diagonalize the specific Hamiltonian.
Using the Green function method, we can include impurity effects and extend them to many-body systems.

\section{\label{sec:summary}Summary}
We established the linear response theory of the spin Nernst effect by using Luttinger's gravitational potential method and considering the contribution from the local equilibrium current.
We clarified that the contribution precisely removes the unphysical divergence of the non-equilibrium spin current at $T=0$ with the Ward-Takahasi identities.
As a demonstration, the spin Nernst effect of three-dimensional Dirac electrons is computed.
Our theory is general and thus straightforward to be extended to interacting electron systems, where Mott's formula no longer holds.

%
\begin{acknowledgments}
The authors thank M.~Sato for the fruitful discussion and T.~Yamaguchi and S.~Fujimoto for their valuable comments.
This work is supported by JSPS KAKENHI Grant No.~JP22K13997 from MEXT, Japan, and JST CREST Grant No.~JPMJCR19T2.
\end{acknowledgments}
%
%
\appendix
\section{\label{apx:cleanlimit}Non-equilibrium and equilibrium current in clean limit}
Here, we show the calculation details in the clean limit.
First, we calculate $F_{1, ij}^{\alpha}$.
In the clean limit, the vertex of velocity of the spin current is reduced to the bare one; $\Lambda_{\mathrm{s}, i}^{\Y \X, \alpha} (\bk; \epsilon', \epsilon) = v_{\mathrm{s}, i}^{\alpha} (\bk)$.
\begin{align}
& F_{1, ij}^{\alpha} (\epsilon + \mu)
\notag \\
    & = -\frac{1}{2 V} \sum_{\bk} \tr \left[
            v_{\mathrm{s}, i}^{\alpha} (\bk)
            \left(G_{\bk}^{\R} - G_{\bk}^{\A} \right)
            v_j (\bk)
            \left(G_{\bk}^{\R} - G_{\bk}^{\A} \right)
        \right]
\notag \\ & \hspace{1em}
        + \frac{1}{2 V} \sum_{\bk} \tr \left[
            v_{\mathrm{s}, i}^{\alpha} (\bk) G^{\R}_{\bk} v_j (\bk) G^{\A}_{\bk}
            - v_{\mathrm{s}, i}^{\alpha} (\bk) G^{\A}_{\bk} v_j (\bk) G^{\R}_{\bk}
        \right]
\label{eq:F1_clean-limit}
,\end{align}
where $G^{\R/\A}_{\bk} = G^{\R/\A}_{\bk} (\epsilon)$.
The first term in Eq.~(\ref{eq:F1_clean-limit}) contributes only to the longitudinal response and the second contributes to the transverse response, so that we omit the first and consider the second term below.
By using Eq.~(\ref{eq:expansion_G}), we have
\begin{align}
& F_{1, ij}^{\alpha} (\epsilon + \mu)
\notag \\
    & =  \frac{1}{2 V} \sum_{\bk} \sum_{n \neq m} \left[
            \frac{\bra{n} v_{\mathrm{s}, i}^{\alpha} (\bk) \ket{m} \bra{m} v_{j} (\bk) \ket{n} }{(\epsilon + \mu - \epsilon_{m \bk} + \zi 0) (\epsilon + \mu - \epsilon_{n \bk} - \zi 0)}
\right. \notag \\ & \hspace{6em} \left.
            - \frac{\bra{m} v_{\mathrm{s}, i}^{\alpha} (\bk) \ket{n} \bra{n} v_{j} (\bk) \ket{m} }{(\epsilon + \mu - \epsilon_{n \bk} - \zi 0) (\epsilon + \mu - \epsilon_{m \bk} + \zi 0)}
        \right]
.\end{align}
For the $\epsilon$-integral in Eq.~(\ref{eq:KuboLuttingerFinalL}), we rewrite the contour integral and take the pole of $\epsilon = \epsilon_{n \bk} - \mu + \zi 0$, and then we get
\begin{align}
& \frac{\hbar^2}{4 \pi} \int_{-\infty}^{\infty} \dd{\epsilon} \left( - \frac{\partial f}{\partial \epsilon} \right) \epsilon F_{1, ij}^{\alpha} (\epsilon + \mu)
\notag \\ & \hspace{4em}
    = - \frac{1}{2 V} \sum_{\bk} \sum_{n} f'_{\mathrm{FD}} (\epsilon_{n \bk}) (\epsilon_{n \bk} - \mu) m_{n \bk, i j}^{\alpha}
,\end{align}
where $f'_{\mathrm{FD}} (\epsilon) = \partial f_{\mathrm{FD}} (\epsilon) / \partial \epsilon$ and $m_{n \bk, i j}^{\alpha}$ is the spin-dependent magnetic moment defined by Eq.~(\ref{eq:m_nk}).

Next, we calculate $F_{2, ij}^{\alpha}$ as
\begin{align}
F_{2, ij}^{\alpha} (\epsilon + \mu)
    & = - \frac{1}{2 V} \sum_{\bk} \tr \left[
        v_{\mathrm{s}, i}^{\alpha} (\bk)
        \left(\partial_{\epsilon} G^{\R}_{\bk} \right)
        v_j (\bk)
        G^{\R}_{\bk}
\right. \notag \\ & \hspace{2em} \left.
        - v_{\mathrm{s}, i}^{\alpha} (\bk)
        G^{\R}_{\bk}
        v_j (\bk)
        \left(\partial_{\epsilon} G^{\R}_{\bk} \right)
        - (\R \to \A)
    \right]
.\end{align}
Here, the derivative of the Green function is calculated as
\begin{align}
\partial_{\epsilon} G^{\R/\A}_{\bk}
    & = \frac{\partial}{\partial \epsilon} \left( \sum_n \frac{\ket{n \bk} \bra{n \bk}}{\epsilon + \mu - \epsilon_{n \bk} \pm \zi 0} \right)
\notag \\
    & = - \sum_n \frac{\ket{n \bk} \bra{n \bk}}{(\epsilon + \mu - \epsilon_{n \bk} \pm \zi 0)^2}
,\end{align}
so that we have
\begin{align}
& F_{2, ij}^{\alpha} (\epsilon + \mu)
\notag \\ &
    = \frac{1}{2 V} \sum_{\bk} \sum_{n \neq m} \left[
        \frac{ \bra{n} v_{\mathrm{s}, i}^{\alpha} (\bk) \ket{m} \bra{m} v_{j} (\bk) \ket{n} }{(\epsilon + \mu - \epsilon_{m \bk} + \zi 0)^2 (\epsilon + \mu - \epsilon_{n \bk} + \zi 0)}
\right. \notag \\ & \hspace{1em} \left.
        - \frac{ \bra{m} v_{\mathrm{s}, i}^{\alpha} (\bk) \ket{n} \bra{n} v_{j} (\bk) \ket{m} }{(\epsilon + \mu - \epsilon_{m \bk} + \zi 0)^2 (\epsilon + \mu - \epsilon_{n \bk} + \zi 0)}
        - (+ \zi 0 \to - \zi 0)
    \right]
.\end{align}
We rewrite the $\epsilon$-integral as the contour integral and we use the residue theorem for $\epsilon = \epsilon_{n \bk} - \mu - \zi 0$ and $\epsilon = \epsilon_{m \bk} - \mu - \zi 0$.
Then, we get
\begin{widetext}
\begin{align}
\frac{\hbar^2}{4 \pi} \int_{-\infty}^{\infty} \dd{\epsilon} f (\epsilon) \epsilon F_{2, ij}^{\alpha} (\epsilon + \mu)
    & = - \frac{\zi \hbar^2}{2 V} \sum_{\bk} \sum_{n \neq m} \left[
            \frac{f_{\mathrm{FD}}(\epsilon_{n \bk}) (\epsilon_{n \bk}-\mu)}{(\epsilon_{n \bk} - \epsilon_{m \bk})^2}
            + \frac{\partial}{\partial \epsilon} \left. \left( \frac{\epsilon f (\epsilon)}{\epsilon + \mu - \epsilon_{n \bk} + \zi 0} \right) \right|_{\epsilon = \epsilon_{m \bk} - \mu - \zi 0}
        \right]
\notag \\ & \hspace{1em}
    \times \left\{
        \bra{n} v_{\mathrm{s}, i}^{\alpha} (\bk) \ket{m} \bra{m} v_{j} (\bk) \ket{n}
        - \bra{m} v_{\mathrm{s}, i}^{\alpha} (\bk) \ket{n} \bra{n} v_{j} (\bk) \ket{m}
    \right\}
\notag \\
    & = - \frac{\zi \hbar^2}{2 V} \sum_{\bk} \sum_{n \neq m} \left[
            2 f_{\mathrm{FD}}(\epsilon_{n \bk}) \frac{\epsilon_{n \bk} - \mu}{(\epsilon_{n \bk} - \epsilon_{m \bk})^2}
            - \frac{f_{\mathrm{FD}} (\epsilon_{n \bk})}{\epsilon_{n \bk} - \epsilon_{m \bk}}
            - f'_{\mathrm{FD}} (\epsilon_{n \bk}) \frac{\epsilon_{n \bk} - \mu}{\epsilon_{n \bk} - \epsilon_{m \bk}}
        \right]
\notag \\ & \hspace{1em}
    \times \left\{
        \bra{n} v_{\mathrm{s}, i}^{\alpha} (\bk) \ket{m} \bra{m} v_{j} (\bk) \ket{n}
        - \bra{m} v_{\mathrm{s}, i}^{\alpha} (\bk) \ket{n} \bra{n} v_{j} (\bk) \ket{m}
    \right\}
\notag \\
    & = - \frac{1}{V} \sum_{\bk} \sum_{n} \left\{
        f_{\mathrm{FD}} (\epsilon_{n \bk}) (\epsilon_{n \bk} - \mu) \Omega_{n \bk, ij}^{\alpha}
        - \frac{1}{2} f_{\mathrm{FD}} (\epsilon_{n \bk}) m_{n \bk, ij}^{\alpha}
        - \frac{1}{2} f'_{\mathrm{FD}} (\epsilon_{n \bk}) (\epsilon_{n \bk} - \mu) m_{n \bk, ij}^{\alpha}
    \right\}
.\end{align}
Finally, we obtain $\tilde{L}_{ij}^{\alpha}$ as
\begin{align}
\tilde{L}_{i j}^{\alpha}
    & = - \frac{1}{V} \sum_{n, \bk} f_{\mathrm{FD}} (\epsilon_{n \bk}) \left(
        (\epsilon_{n \bk} - \mu) \Omega_{n \bk, ij}^{\alpha}
        - \frac{1}{2} m_{n \bk, ij}^{\alpha}
    \right)
,\end{align}
which is Eq.~(\ref{eq:clean_limit_L_tilde}) in the main text.

Similarly, for the equilibrium current, we calculate
\begin{align}
\mathcal{J}_{ij}^{0, \alpha} [\mu, T] & =
\frac{\hbar^2 \kB T}{4 \pi} \int_{-\infty}^{\infty} \dd{\epsilon} F_{2, ij}^{\alpha} (\epsilon)\ln 
         \left[ 1 + \exp \left( - \frac{\epsilon - \mu}{\kB T} \right) \right] \notag \\ 
    & = - \frac{\zi \hbar^2 \kB T}{2 V} \sum_{\bk} \sum_{n \neq m} \left[
            \frac{\ln \left[1 + \exp \left( - \frac{\epsilon_{n \bk} - \mu}{\kB T} \right) \right] }{(\epsilon_{n \bk} - \epsilon_{m \bk})^2}
            + \frac{\partial}{\partial \epsilon} \left. \left( \frac{\ln \left[ 1 + \exp \left( - \frac{\epsilon - \mu}{\kB T} \right) \right] }{\epsilon + \mu - \epsilon_{n \bk} + \zi 0} \right) \right|_{\epsilon = \epsilon_{m \bk} - \mu - \zi 0}
        \right]
\notag \\ & \hspace{1em}
    \times \left\{
        \bra{n} v_{\mathrm{s}, i}^{\alpha} (\bk) \ket{m} \bra{m} v_{j} (\bk) \ket{n}
        - \bra{m} v_{\mathrm{s}, i}^{\alpha} (\bk) \ket{n} \bra{n} v_{j} (\bk) \ket{m}
    \right\}
\notag \\
    & = - \frac{1}{V} \sum_{\bk} \sum_{n} \left\{
        \kB T \Omega_{n \bk, ij}^{\alpha} \ln \left[1 + \exp \left( - \frac{\epsilon_{n \bk} - \mu}{\kB T} \right) \right]
        + \frac{1}{2} f_{\mathrm{FD}} (\epsilon_{n \bk}) m_{n \bk, ij}^{\alpha}
    \right\}
,\end{align}
which is Eq.~(\ref{eq:clean_limit_J0}) in the main text.
\end{widetext}

\bibliography{spin-Nernst}
\end{document}